# Information processing at the speed of light

**Muhammad AbuGhanem**[1]

**Abstract**
In recent years, quantum computing has made significant strides, particularly in light-based technology. The introduction of quantum photonic chips has ushered in an era marked by scalability, stability, and cost-effectiveness, paving the way for innovative possibilities within compact footprints. This article provides a comprehensive exploration of photonic quantum computing, covering key aspects such as encoding information in photons, the merits of photonic qubits, and essential photonic device components including light squeezers, quantum light sources, interferometers, photodetectors, and waveguides. The article also examines photonic quantum communication and internet, and its implications for secure systems, detailing implementations such as quantum key distribution and long-distance communication. Emerging trends in quantum communication and essential reconfigurable elements for advancing photonic quantum internet are discussed. The review further navigates the path towards establishing scalable and fault-tolerant photonic quantum computers, highlighting quantum computational advantages achieved using photons. Additionally, the discussion extends to programmable photonic circuits, integrated photonics and transformative applications. Lastly, the review addresses prospects, implications, and challenges in photonic quantum computing, offering valuable insights into current advancements and promising future directions in this technology.

**Keywords** Photonics quantum computing · Nobel Prize-winning technology · Integrated photonics · Photonic device components · Encoding information in photons · Programmable photonic circuits · Photonic quantum computers · Quantum communication and internet · Quantum key distribution · Free-space communication · Quantum computational advantage with photons

## 1 Introduction

Since its inception a century ago, quantum mechanics has continually presented unexpected discoveries, with significant breakthroughs occurring nearly every year. In recent decades, there has been a notable surge in both theoretical and practical advancements in this field [1–19]. Among the most captivating frontiers within quantum mechanics is the pursuit of genuine quantum computers [1, 2, 20–25], machines poised to tackle computational tasks far beyond the capabilities of classical computing counterparts [26–33].

Light-based quantum technologies [34–36] currently stand as prominent candidates in the domain of fault-tolerant quantum computation (FTQC) [37]. These sophisticated architectures, delineating a novel paradigm for quantum information processing (QIP), employ photons as the medium for qubit encoding and manipulation [38]. Notably, they demonstrate intrinsic resilience against decoherence and noise, even at room temperature, rendering them exceptionally suitable for scalable and FTQC [37].

Photonic quantum computing [39, 40] represents a distinctive approach with a unique set of advantages that sets it apart from other quantum computing methodologies. Significantly, photonics has emerged as the exclusive platform capable of constructing modular, easily-networked quantum computers operating at room temperature, holding substantial promise for practical quantum applications [1]. A key attribute of photonic quantum computing lies in the encoding of qubits within the quantum state of light, unlocking a plethora of possibilities for QIP. Quantum states of light have played a pivotal role since the inception of groundbreaking experiments in nonlocality and quantum teleportation [39, 40].

✉ Muhammad AbuGhanem
gaa1nem@gmail.com

[1] Faculty of Science, Ain Shams University, Cairo 11566, Egypt

Over the past decade, remarkable advancements in photonic quantum technologies [36] have led to groundbreaking achievements in various quantum information science domains. Notable milestones include the establishment of satellite quantum communications (QCOMM) [41, 42] and the realization of quantum computational advantage (QCOA) [43–45]. Recently, photonic processors have garnered significant interest due to their diverse applications, spanning QIP based on linear optics [46–58], quantum machine learning and artificial intelligence [59–63], radio-frequency signal processing [64, 65], and quantum repeater networks [66–69]. Operating as tunable multimode interferometers capable of executing arbitrary linear optical transformations, photonic processors have been manifested in various topologies, including triangular [46, 70], square [71], fan-like [60], hexagonal [65], quadratic [64], and rhomboidal configurations [49]. These applications showcase the versatility and potential impact of photonic quantum technologies in advancing quantum information science. In this paper, we offer an in-depth overview of pivotal facets within the field of photonic quantum computing.

The rest of this paper is structured as follows: Sect. 2 delves into light-based QIP, laying the foundation for subsequent discussions. Section 3 explores diverse approaches for encoding information in photons. The advantages and disadvantages of photonic qubits are discussed in Sect. 4, followed by a detailed examination of GKP-encoded states and Gaussian channels in Sect. 5. Section 6 outlines the photonic technology required for building photonic quantum computers, including light squeezers, quantum light sources, interferometers, photodetectors, waveguide and linear optical networks (LONs). Section 7 outlines the trajectory towards architecting scalable and fault-tolerant quantum computing using photonic technologies. Section 8 elucidates programmable photonic circuits, while Sect. 9 provide an overview of quantum photonic communication and internet. The section also discusses: the reconfigurable photonic technology for QCOMM 9.1, secure QCOMM systems 9.2, implementations of photonic quantum key distribution (QKD) 9.3, long-distance QCOMM 9.4, QKD performance parameters 9.5, and QCOMM trends in research and patents 9.6. Section 10 explores the QCOAs witnessed in specific cases, whereas Sect. 11 assesses practical applications of photonic quantum computers. The novel prospects, implications, and challenges in photonic quantum computing are discussed in Sect. 12. Finally, our conclusions are drawn in Sect. 13.

## 2 Photonic quantum computing

Since the inception of quantum computing, optical quantum systems [38] have consistently maintained a prominent standing as primary candidates for its realization [72]. These systems leverage the well-established framework of quantum optics to manipulate quantum states of light, thereby facilitating the execution of quantum computations [38].

Photons have emerged as a flagship system for delving into the intricacies of quantum mechanics, propelling advancements in quantum information science (QIS), and catalyzing the evolution of quantum technologies [39, 40]. Pioneering breakthroughs in quantum entanglement [109], teleportation, QKD, and early quantum computing demonstrations have been achieved through the utilization of photons [18, 110–114]. This preference arises from photons serving as a naturally mobile and low-noise system, complemented by the availability of quantum-limited detection mechanisms [114]. The quantum states of individual photons can be precisely manipulated using interferometry. The sophistication of photonic quantum computing devices and the realization of protocols have surged forward, propelled by ongoing developments in both underlying technologies and theoretical frameworks [40].

Recently, photonic quantum computing stands as a compelling avenue for achieving medium- and large-scale processing capabilities, overcoming historical associations with resource-intensive requirements linked to inefficient two-qubit gates [115–117]. The adept generation of abundant photons, coupled with advancements in integrated platforms, enhanced sources and detectors, innovative noise-tolerant theoretical approaches, and more, firmly establishes photonic quantum computing as a prominent candidate for both QIP and quantum networking [40].

Figure 1 summarizes the key milestones and advancements in photonic quantum computing research over the past 15 years, from 2009 to 2023 [73, 74]. These progressions are discussed in details in the subsequent sections, particularly Sect. 9, Sect. 10, and Sect. 11. Milestones include the first on-chip quantum interference and integrated controlled-NOT (CNOT) gate [118] (not depicted), testing of Shor's factoring algorithm [75], laser-writing of integrated quantum photonic circuits [76], and the first re-programmable multi-functional quantum processor unit (QPU) [57], among others [1]. These advancements encompass waveguide detectors with photon number resolution (PNR), efficient single-photon sources from quantum dots (QD) [86], demonstrations of Boson sampling (BS) with multiple photons [55, 56, 82–84], and the realization of large-scale quantum circuits [97, 107]. Recent highlights include quantum supremacy experiments (QSEs) [43–45, 99, 108] and advancements in QCOMM networks and processors [103], culminating in 2023's achievements in BS and pseudo-photon-number-resolving detection [108]. Note that specific years on the figure may differ from those in the references, since the chronological order is based on the earliest appearances in public sources, such as conferences and preprints, as adapted from [73, 74].

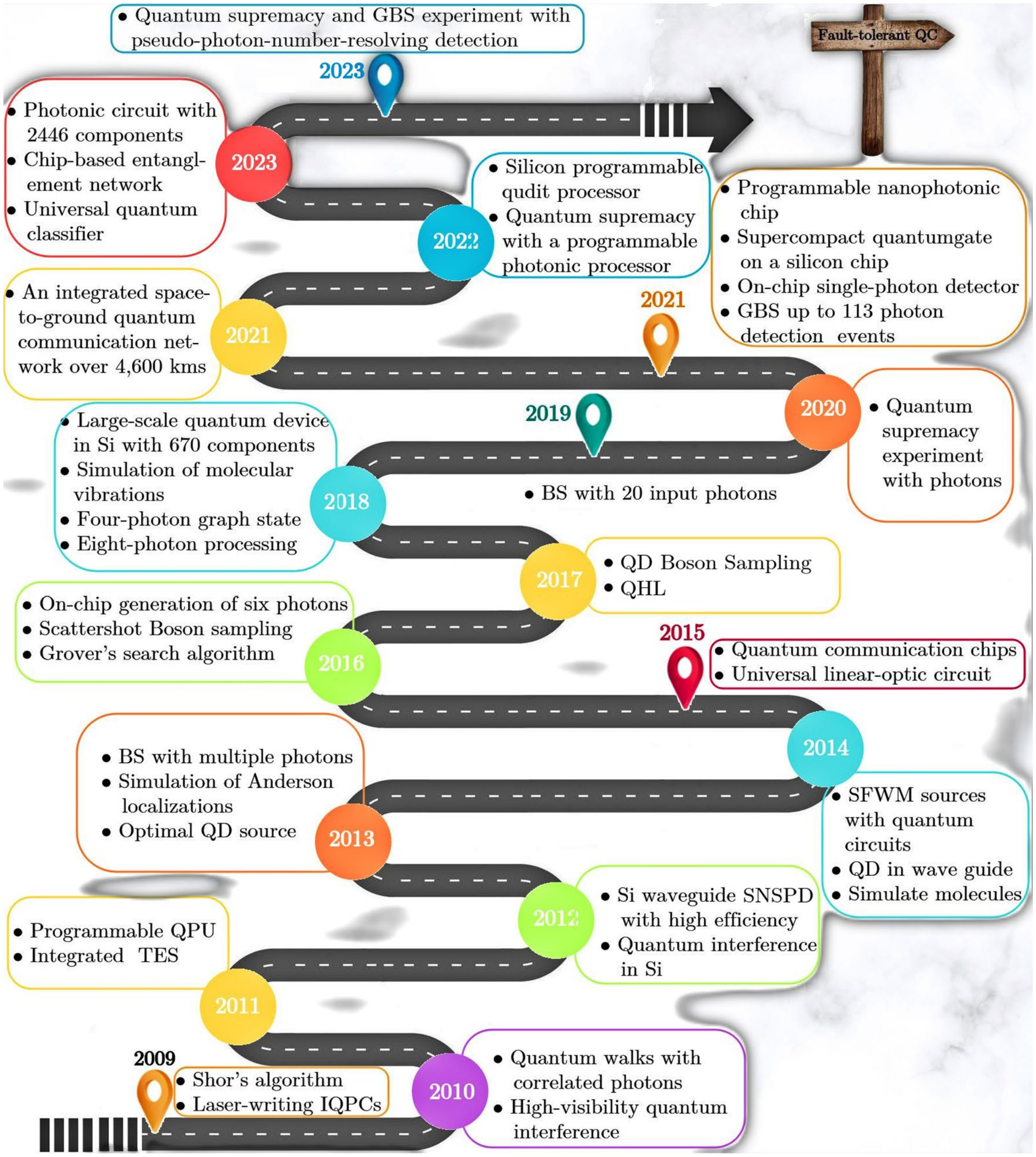

**Fig. 1** Progression and key milestones in integrated quantum photonics over the past 15 years (2009–2023) [1, 73, 74]. Milestones include testing of Shor's algorithm [75], laser-writing integrated quantum photonic circuits [76], quantum walks [77], near optimal 2-photon quantum interference [78], the 1st programmable QPU [57], a waveguide transition edge sensor (TES) with photon-number-resolving (PNR) capability [79], Si waveguide SNSPDs (superconducting nanowire single photon detectors) [80]. quantum interference in Si [81], Boson sampling (BS) experiments utilizing multiple photons [55, 56, 82–84], simulating Anderson localizations via entangled photons [85], near-optimal single-photon generation from a single quantum dot (QD) [86], the earliest integration of spontaneous four-wave mixing (SFWM) sources with quantum circuits [87], coupling QD sources into waveguides [88], simulating molecular ground states [89], chip-to-chip QKD and entanglement distribution [90, 91], universal linear-optic circuit [46], on-chip generation of 6 photons [92], quantum circuits of scattershot BS [93], testing Grover's search algorithm [94], high-efficiency BS using with a QD source [95], test of QHL (quantum Hamiltonian learning) algorithm [96], quantum circuits in Si with 670 components [97], simulating molecular vibrations [53], four-photon graph state demonstration [98], on-chip generation of 8 photons in Si [58], BS with 20 input photons [99], quantum supremacy experiment (QSE) with photons [43], programmable chip [100], supercompact photonic quantum gate [101], on-chip single-photon detector [102], Gaussian BS up to 113 photon detection events [44], an integrated space-to-ground QCOMM network over 4,600 kms [103], a programmable qudit-based processor [104], QSE with a programmable photonic processor [45], multi-chip multi-dimentional quantum network [105], demonstration of bosonic quantum classifier [106], photonic circuit with 2446 components [107], and QSE and GBS with pseudo-photon-number-resolving detection [108]

Linear optics QIP stands as a promising methodology poised to tackle computational challenges that far exceed the capabilities of classical computers exponentially [35]. This approach has encouraged the proposal of numerous applications spanning various domains [1, 35, 36, 119]. The recent validation of QCOA [1] within a static optical system underscores the compelling need for programmable photonic processors [43].

At the heart of linear optics QIP lies the foundational mechanism of quantum interference. This intricate process involves an arrangement comprising photon sources, a photonic processor, and single-photon detectors (as detailed in Sect. 6). Photons, serving as carriers of information, undergo manipulation by the photonic processor, which is composed of linear optical elements. This manipulation involves the orchestrated control of interference among photons. The computational outcome of the photonic system can be precisely determined by scrutinizing the configurations of output samples derived from the detected photons [120–126].

The concept of LOQC (linear optical quantum computing), grounded in elementary yet probabilistic quantum operations, has garnered growing optimism and has evolved steadily over the past two decades. Notable advancements in LOQC highlight its increasing promise for practical QIP. A deeper exploration of the historical evolution of this field is available in [24, 34–36]. These reviews offer valuable insights into the earlier stages of LOQC development, paving the way for a more nuanced understanding of its current landscape.

The effective realization of photonic QIP demands that photonic processors adhere to four fundamental criteria. *Firstly,* they must boast a large-scale architecture capable of handling complex problem-solving tasks. *Secondly,* universality is imperative, facilitating the implementation of arbitrary transformations that map the system onto diverse problem domains. The attainment of universality necessitates both all-to-all connectivity and full reconfigurability [46, 70, 71, 127]. *Thirdly,* maintaining low loss is of paramount importance to preserve the integrity of (quantum) information carried by the system. *Lastly,* a photonic processor must effectively uphold quantum interference, ensuring the precision and reliability of quantum computations [128].

Photonic platforms naturally offer certain experimental advantages over alternative platforms. Notably, quantum information is conventionally encoded within photons. Photons exhibit limited interactions both amongst themselves and with their environment, endowing them with a notable resistance to the challenges posed by decoherence. However, this inherent virtue also engenders a challenge, as managing interactions among individual photons proves to be a complex endeavor, thereby complicating the realization of two-qubit quantum gates. Initial proposals for introducing photon interactions revolved around two predominant strategies: representing $k$ qubits by means of a single photon traversing $2^k$ distinct modes or pathways [129], and harnessing nonlinear elements, such as a Kerr medium [130]. Unfortunately, both of these approaches encountered substantial limitations - the former in terms of scalability and the latter due to the formidable experimental intricacies [131].

In the year 2001, KLM (Knill, Laflamme, and Milburn) [18] elucidated that the realization of universal quantum computing on a photonic platform could be theoretically achieved employing a minimalistic set of components, namely beam splitters (B-Ss), phase shifters (PSs), single photon sources (SPSs), and photon detectors (PDs). This discovery is of paramount significance, given that it circumvents the necessity for non-linear interactions between optical modes, except potentially during the initial state preparation, rendering it considerably more amenable for practical implementation. It is conceivable to induce nonlinearity through post-selection, although this approach renders the scheme probabilistic. The protocol, famously known as the KLM protocol, also underscores that such a platform necessitates an exponential abundance of resources to surmount the inherent probabilistic nature of linear optics [132].

Light-based quantum architectures continue to grapple with substantial scalability issues, predominantly arising from the inherent probabilistic nature of the KLM protocol. The KLM protocol is renowned for addressing the exponential resource requirements needed to manage this probabilistic behavior in linear optics. This probabilistic nature poses challenges for scalability, particularly in achieving deterministic production of single photons. The reliable generation and assembly of these photons into more complex quantum states at scale prove challenging [35]. One proposed alternative is the utilization of Gaussian states, specifically squeezed states of light that do not comprise single photons, offering a considerably higher degree of experimental controllability [133]. This approach is closely aligned with continuous variable quantum computation, a quantum computing paradigm that operates within infinite-dimensional Hilbert spaces rather than finite-dimensional Hilbert spaces associated with qubits [134–136].

KLM have proposed an effective framework for quantum computation utilizing linear optics [18]. Numerous investigations have been conducted in the context of an intense laser regime to explore the implementation of diverse protocols, including quantum gates [137], optical communication [138], QKD [139], quantum channels [140], teleportation [141], as well as the examination of decoherence effects in both Markovian and non-Markovian evolutions [142], the development of quantum thermal engines [143], and environment-induced entanglement [144].

In recent years, there has been a notable development focused on the establishment of protocols designed to authenticate the production of pure 2-qubit states [145–149]. In experimental settings, the Mach-Zehnder and Sagnac interferometers are frequently employed for the preparation and measurement of two-qubit states [150], as well as for implementing 2-qubit quantum logic gates [151]. Recently, Gonzales et al. [150] have introduced and validated a methodology for generating and characterizing pure 2-qubit states utilizing a Mach-Zehnder-type interferometer (MZI). Practical implementation of the generation procedure outlined in [150] necessitates meticulous precision for the preparation of states featuring arbitrary phases. Additionally, a procedural framework for the synthesis and characterization of pure two-qubit states, wherein the encoding involves the polarization and momentum (path) attributes of light beams, has been explored [152].

It is worth emphasizing that photons serve as a highly effective platform for QCOMMs [153–155]. For further in-depth exploration of photonic quantum computing, see [35, 133, 156]. An insightful discussion on experimental approaches toward constructing a scalable photonic quantum computer is available in [157]. Also, a review on optomechanics for quantum technologies was presented in [158], focusing on opto- and electromechanical platforms [159].

## 3 Encoding information in photons

Light-based QIP leverages the inherent degrees of freedom (DoFs) within light, encompassing characteristics associated with spin angular momentum (polarization encoding), propagation directions (path encoding), spatial distribution of light (orbital angular momentum encoding), and time encoding (time-bin and time-frequency encoding). Each encoding strategy presents unique advantages and limitations, which can be tactfully combined in a hybrid configuration to optimize performance and functionality.

Photons emerge as highly effective carriers of quantum information, thanks to their prolonged coherence times at room temperature, rendering them indispensable for disseminating quantum information across extensive distances, whether through free space or optical fiber networks [103, 160, 161]. A critical aspect of utilizing photonic qubits lies in the precise initialization of quantum states. This initial state preparation is pivotal as subsequent adjustments to entanglement after emission proves nontrivial [162]. Initialization strategies are contingent upon the chosen DoFs for encoding quantum information, and within the domain of QCOMM over optical channels, time-bin encoding [163, 164] stands out as the prevailing choice, wherein the two-qubit levels denote the photon occupying one of two time windows typically separated by a few nanoseconds. Time-bin encoding exhibits exceptional resilience against phase fluctuations arising from thermal noise in optical fibers, maintaining qubit coherence even over distances spanning hundreds of kilometers [165, 166].

Despite its advantages, manipulating the state of time-bin-entangled photons faces challenges, especially in the context of emerging nano-photonic platforms. For on-chip qubit state manipulation, the dual-rail encoding approach proves superior [35, 167], wherein the two states of a qubit correspond to the photon propagating in one of two optical waveguides. This strategy is commonly preferred for quantum computing and quantum simulation in integrated platforms. However, it poses compatibility challenges for long-distance transmission links, whether through optical fibers or free space channels. The field of photonic quantum computing offers a rich assortment of methodologies for encoding quantum information in photons. In this section, we explore distinct representations of photonic qubits and their applications in the realm of QIP.

### 3.1 Polarization-based qubits

Polarization stands as a fundamental avenue for qubit encoding within photonic systems [168]. By utilizing the two orthogonal polarizations of the electromagnetic field (e.g. horizontal and vertical), this method finds substantial utility, notably in the domain of LOQC [35, 169]. The well-established framework of polarization-based qubits underscores their potential for robust qubit manipulation in quantum computations [8, 36, 170, 171]. The capacity to manipulate quantum states of light through integrated devices holds significant potential for conducting fundamental assessments of quantum mechanics and exploring innovative technological applications [170]. The first demonstration of an integrated photonic CNOT gate tailored for polarization-encoded qubits is presented in [172]. This achievement was facilitated through integration methodologies grounded in femtosecond laser waveguide writing [173–177], wherein partially polarizing B-Ss were incorporated onto a glass chip [172].

In polarization encoding, a quantum bit typically appears as $|\Omega\rangle = \lambda|V\rangle + \gamma|H\rangle$, with $V$ and $H$ denoting vertical and horizontal polarization respectively. Additionally, the expression

$$|p\rangle = \int_{-\infty}^{\infty} \mathrm{d}\mathbf{k}\, y(\mathbf{k})\, \mathrm{e}^{-\mathrm{i}\omega_k t}\, \hat{a}^{\dagger}(\mathbf{k}, p)|0\rangle, \tag{1}$$

describes the state, where $p = (H, V)$, $y(\cdot)$ represents a wave packet mode function, and $\hat{a}^{\dagger}(k, p)$ stands for the creation operator of a photon with momentum $\mathbf{k}$ and polarization $p$ [39, 178]. Polarization qubits can also be represented in alternative bases, such as the left and right circular polarizations as $|L/R\rangle = (|H\rangle \pm \mathrm{i}|V\rangle)/\sqrt{2}$, or with the diagonal basis $|\pm\rangle = (|H\rangle \pm \mathrm{i}|V\rangle)/\sqrt{2}$. These three pairs of states collectively form a set of mutually unbiased bases encoded in polarization, which are fundamental to numerous applications [179].

Polarization encoding has long been crucial in various quantum information investigations, spanning from quantum computation [8, 94, 180–182] to simulation [85, 183] and communication [184, 185]. Its widespread utilization in QIP has been bolstered by advancements in entanglement generation, manipulation, and distribution [90, 169, 186–196]. Furthermore, polarization qubits are increasingly interconnected with other photon DoFs, such as path encoding [8, 90, 94, 180, 197, 198], orbital angular momentum [191, 196, 199, 200], time-energy [201, 202], and their combinations in hyper-entangled states [111, 203]. These states serve as efficient resources for protocols in quantum computation and communication [39].

### 3.2 Path-encoded qubits

Path encoding emerges as an intriguing approach for qubit representation, relying on the distinction between the two transmission paths of single photons. While concerns about phase stability may arise in free space, this approach demonstrates notable suitability for integrated photonics. The path-based qubits open possibilities for scalable and compact quantum computing architectures [204]. In integrated photonics, qubits find encoding in the path DoFs, commonly referred to as "dual-rail" qubits or simply path qubits. A path qubit manifests as a photon distributed across two spatial modes, specifically within the confines of single-mode waveguides. Consequently, a path qubit is situated within a pair of waveguides. Through the encoding of qubits in the DoFs of light, the construction of a two-qubit quantum gate using linear optical devices becomes feasible [205–207]. Consequently, numerous studies underscore the advantages associated with leveraging the DoFs of light as a supplementary resource for various aspects of quantum teleportation [191], quantum computation [208, 209], communication [210], and quantum cryptography [211].

Path-encoded qubits are well-suited for photonic integrated circuits [57, 94, 212], benefiting from spatial separation in waveguide arrays and easy mode coupling through directional couplers. Experimental work has demonstrated tunable all-optical path entanglement, primarily on SoI and $LiNbO_3$ platforms [87, 213–216]. Fiber-integrated sources for high-dimensional qubits [217], with investigations into mitigating non-ideal implementations through loss analysis and state tomographies [57, 218].

### 3.3 Time-bin encoding

Qubits can be effectively encoded by analyzing the arrival times of single photons, distinguishing between early and late arrivals. The time bin quantum states offer a practical avenue for encoding quantum information and exhibit relevance in specific QCOMM protocols and photonic quantum gates [163, 164]. Time-bin qubits were initially formulated by Brendel et al. [163], who illustrated their capability to traverse extended distances within optical fibers with minimal decoherence, thus presenting a more resilient foundation for QCOMM systems in comparison to those reliant on polarization-encoded qubits [165, 219]. Humphreys et al. [220] pioneered an optical quantum computing methodology utilizing time-bin qubits. Additionally, Donohue et al. [221] validated an expeditious measurement technique for time-bin qubits, holding promise for heightened data rates and reduced errors in photonic systems. Expanding the application of time-bin qubits beyond QCOMM, Humphreys et al. demonstrated the feasibility of employing them for quantum computing [220]. Their approach aligns with the established paradigm of LOQC, wherein abundant ancilla photons and measurement-based nonlinearities are harnessed to achieve nearly deterministic quantum logic gates [18].

Ortu et al. [222] successfully achieved the storage of six distinct temporal modes for durations of 20, 50, and 100 ms within a rare-earth doped crystal, specifically the $^{151}Eu^{3+} : Y_2SiO_5$ crystal. The quantum coherence of the implemented memory system was substantiated through the storage of two time-bin qubits over a 20 ms interval, yielding an average memory output fidelity of $\mathcal{F} = (85 \pm 2)\%$, considering an average number of photons per qubit of $\mu_{\mathrm{in}} = 0.92 \pm 0.04$. In [223], an up-conversion single-photon detector (UCSPD) was innovatively engineered by employing commercial nonlinear crystals of varying lengths. The study effectively assessed pseudo femtosecond time-bin qubits at the single-photon level, achieving a remarkable pulse interval of merely 800 fs. Furthermore, the experimental validation of the viability of picosecond time-bin states of light, denoted as ultrafast time-bins, for utilization in QCOMMs was successfully demonstrated [224].

The method for encoding information utilizes an MZI with one arm longer than the other. When an incoming photon's amplitude is split at the first B-S of the Mach-Zehnder, it traverses through the unbalanced arms: we designate the state of a photon that has traveled the long path as $|l\rangle$, while $|s\rangle$ denotes a photon that has taken the shorter path. Thus, in the photon arrival time, a qubit encoded can be expressed as the superposition $|\Omega\rangle = \lambda|l\rangle + \gamma|s\rangle$, where the states $|p\rangle$ are defined as

$$|p\rangle = \int_{-\infty}^{\infty} \mathrm{d}x\, y\Big(\frac{t + p\tau - x/c}{\delta t}\Big) \exp\big(-\mathrm{i}\omega(t - \frac{L}{c} + p\tau)\big)\, \hat{a}^{\dagger}|0\rangle, \tag{2}$$

with $p = (l, s)$, where $\omega$ is a fixed angular frequency, $y(\cdot)$ is a wave packet mode function, and $\tau$ is the time delay occurred between the two arms of length $L$.

Time-bin encoding offers several advantages over alternative platforms. Firstly, it is well-suited for integrated photonic devices, enabling the generation, manipulation, and measurement of photons without the requirement of external encoding apparatus. Additionally, its robustness against noise affecting polarization, such as depolarizing media, decoherence, and mode dispersion, positions time-bin encoding as a promising choice for applications including QKD and QCOMM, both in free-space and within fiber mediums [165, 225–227], and state teleportation [228–230].

Numerous experiments have been conducted to explore its potential for testing non-locality [164, 165, 221, 231], with entangled photon pairs generated using femtosecond pulses [164] and various integrated sources, including atom-cavity setups and waveguide-based systems [232–235]. Development in on-chip time-bin manipulation [236], storage [227], and measurement [221] has provided crucial components for linear-optical quantum networks. Beyond QCOMM, time-bin encoding has been proposed as a viable scheme for photonic BS [237, 238] (see Sect. 10.2), and quantum walks [239–243].

### 3.4 Frequency-bin encoding

Recently, there has been a notable proposal and experimental demonstration of frequency-bin encoding as a compelling strategy that combines the advantageous characteristics of both time-bin and dual-rail encodings [87, 122, 244–247]. In this approach, quantum information is encoded through the photon existing in a superposition of distinct frequency bands. Manipulation of frequency bins is achieved using phase modulators, and these bins exhibit resilience to phase noise during long-distance propagation.

Pioneering studies have explored the generation and manipulation of frequency-bin-entangled photons within integrated resonators, encompassing investigations into quantum state tomography of entangled photon pairs [248], qudit encoding [120], and the creation of multi-photon entangled states [249]. The successful realization of these experimental results has been made possible by the recent advancements in high-quality integrated resonators within the silicon nitride and silicon oxynitride platforms. Recently, a programmable silicon nano-photonic chip capable of generating frequency-bin entangled photons was successfully demonstrated in [162]. This encoding scheme exhibits compatibility with long-range transmission over optical links [162].

Frequency-encoding finds pertinent applications in QCOMM and QKD, leveraging its low decoherence characteristics for efficient qubit delivery [163, 202, 250–254]. This is evidenced by a range of experimental demonstrations in quantum computation [220, 255–261]. Additionally, ongoing investigations into manipulating photonic qubits encoded in time [221, 260, 262] and time-frequency domains [244, 252, 263–267].

### 3.5 Photon-number encoding

Photon number encoding relies on the vacuum and single-photon states to represent qubit values of 0 and 1, respectively. This straightforward and robust approach has implications in QCOMM and photonic gates, simplifying qubit preparation and manipulation. In photon-number encoding, the information is encoded in the number of photons rather than the specific polarization or other properties of individual photons. The quantum states are prepared in such a way that the number of photons in the state carries the information. This approach has applications in QCOMM [268–272], quantum cryptography [273–276], and quantum computing [277–279].

Photon-number encoding is particularly relevant in QKD protocols [280–282] and quantum repeater schemes [282–284]. It allows for the efficient manipulation and transmission of quantum information while overcoming certain challenges associated with other encoding schemes [232, 273, 274, 285]. An advanced protocol for a photon-number encoded MDI repeater [286–288], which exceeds the PLOB bound [160, 268, 288–293], was introduced in [294]. This approach achieves surpassing performance without reliance on quantum memories [295–299], employing instead an entanglement swapping [294, 300–302].

### 3.6 Orbital angular momentum qubits

Orbital angular momentum or OAM-based qubits explore the spatial distribution of light as a foundational concept [303, 304]. The quantization of a photon's OAM as an integer multiple of $\hbar$ (Planck's constant) offers an innovative avenue for qubit encoding [303]. By utilizing different OAM states, photonic qubits are formed, presenting unique opportunities for QIP [305].

The OAM of light [303] stands out among optical DoFs, offering unique properties [306] for light manipulation [307], heightened sensitivity in imaging [308], and potential high-density information coding in optical communication [309]. Recent interest in leveraging

OAM at the single-photon level for quantum information technologies has grown [310, 311]. The helical shape and orthogonality of OAM beams distinguish them significantly, rendering them applicable in various domains, including optical tweezers [312, 313], atomic manipulation [314–316], nanoscale microscopy [317–320], wireless communication [321], information encoding [322], data storage [323–325], QIP [199, 311, 326–330], and optical communication [331, 332].

The OAM of photons has fostered theoretical and experimental efforts for encoding and processing quantum information [306, 333]. Building on pioneering work demonstrating entanglement in OAM [310], significant progress has been made in experimentally controlling OAM state superpositions. These advances encompass various protocols, including quantum cryptography [334], bit commitment [335], experimental quantum coin tossing [336], and the demonstration of high-dimensional entanglement [199, 337]. Beyond their fundamental significance, these experiments highlight the potential of OAM of light as a carrier of quantum information, promising increased information coding density and advanced processing capabilities, extending to practical applications facilitating optimal quantum cloning [338, 339], long-distance QCOMM [196, 208, 211, 340–349] — such as long-distance quantum repeaters (QRs) in quantum networks [350]— and photonic quantum walks [351–356].

It is evident that OAM states play a foundational role in various quantum information applications [357]. These states enable the preparation of 'flying' qubits without the need for alignment, which is crucial for robust QKD and communication [211, 338, 342, 358–364]. Hybrid encoding schemes incorporating polarization enhance sensitivity to angular rotations [365, 366], while ongoing research supports the development of OAM-based photonic networks for tasks such as routing, sorting, teleportation, and quantum memories [298, 341, 367–371]. Initial efforts toward integrating OAM devices have been undertaken [372], with both theoretical and experimental investigations addressing challenges in free-space communication [340, 343, 349, 373–376]. Storing OAM superpositions at the single-photon level in material systems is crucial for future developments [377–384], with notable progress in storage and manipulation of OAM-encoded quantum information in materials [305].

## 4 Photonic qubits: pros and cons

While pursuing the development of a scalable and fault-tolerant quantum computer, photonic technologies offer significant advantages compared to alternative approaches. These advantages encompass:

- **Room-temperature computation:** Photonic technologies permit computations at room temperature, facilitating complete miniaturization, large-scale manufacturing, the utilization of cost-effective off-the-shelf components, enhanced processing speed, and rapid scalability to a substantial number of qubits by harnessing established silicon electronics and photonics technology.
- **Long coherence time:** Photonic qubits have the advantage of long coherence times, easiness in entanglement generation, and the ability to transmit quantum information over long distances using fiber-optic network.
- **Inherent compatibility with communication technology:** Photonic systems inherently align with communication technology, enabling high-fidelity interconnections between multiple modules without necessitating noisy transduction steps typical of other quantum computing platforms.
- **Flexibility in error-correcting codes:** Photonic technologies offer the flexibility to select error-correcting codes, encompassing both mode-to-qubit encodings and high-dimensional qubit codes that exploit the temporal degrees of freedom of light.
- **Comprehensive all-to-all connectivity:** Photonic qubits offer the advantage of facilitating all-to-all connectivity within their systems, enhancing their computational capabilities.
- **Favorable waveguide dimensions:** Photonic waveguides boast relatively larger dimensions and ease of fabrication, simplifying the manufacturing process.
- **Extensive scale:** Photonic quantum computing stands as a pioneering approach in attaining large number of qubits. It is widely regarded as the most promising architecture for realizing the ambitious goal of 1 million qubits [385].

These advantages [386, 387] present a compelling rationale for the in-depth exploration of architectural frameworks for photonic quantum computation, as they hold the potential to significantly impact the development of scalable and FTQCs. Nevertheless, hurdles encompassing these aspects act as constraints, limiting the scalability of photonic quantum computers [387]. Key challenges include:

- **Low tolerance to detector imperfections:** Photonic qubits exhibit limited tolerance to imperfections in detectors, which can pose challenges in achieving error-free quantum computations.
- **Optical loss errors with the difficulty of integrating a substantial number of high-fidelity multi-photon gates into optical circuits:** This complexity limiting the scalability of photonic quantum computers.

- **Non-standard computation method:** The utilization of photonic qubits typically involves non-standard computation methods, particularly following the principles of one-way quantum computing. Furthermore, photonics exhibits minimal inherent interaction between photons, necessitating the utilization of alternative media for the implementation of logic gates.

## 5 GKP-encoded qubits and Gaussian channels

### 5.1 Overview

Quantum computation (QC) holds considerable potential for efficiently addressing challenging problems beyond the capabilities of classical computers [388, 389]. To achieve the realization of large-scale QC, a promising avenue lies in utilizing continuous-variable (CV) systems [37, 390, 391]. Notably, experimental advancements have been made in generating larger-scale cluster states through the use of squeezed vacuum states in optical setups [258, 392–396]. Moreover, optical setups have successfully produced over a thousand frequency-encoded cluster states [253, 259, 397, 398]. Beyond optical configurations, other platforms such as circuit QED [399], opto-mechanics [400, 401], atomic ensembles [402, 403], and trapped ion mechanical oscillators [404, 405] present promising prospects for large-scale QC involving CVs.

In the pursuit of continuous-variable fault-tolerant quantum computing (CV-FTQC), it is recognized that CVs must be encoded using appropriate bosonic codes [406–409], such as the cat code [410], binomial code [411], or the Gottesman-Kitaev-Preskill (GKP) code [412], denoted as the GKP qubit in this context. This choice is imperative due to the inherent limitations of the squeezed vacuum state in handling analog errors, including those arising from the Gaussian quantum channel [412] and photon loss during QC. Menicucci established the threshold for the squeezing level required for CV-FTQC [413], employing the GKP qubit for quantum error correction in measurement-based QC (MBQC). Recent endeavors have focused on CV-FTQC utilizing the GKP qubit [414–432], including the development of a promising architecture for a scalable quantum circuit incorporating this encoding [157, 433, 434]. Additionally, the GKP qubit emerges as a pivotal element in various QIP applications, notably in long-distance QCOMM [435, 436].

In [415], the necessary squeezing level for CV-FTQC has been reduced to below 10 dB, a threshold achievable with current experimental technology [437]. Notably, in both ion trap setups [405] and superconducting circuit quantum electrodynamics [438], the GKP qubit has been recently generated with squeezing levels approaching 10 dB. However, in optical configurations, despite extensive efforts to explore diverse methods for GKP qubit generation [403, 439–450], the optical GKP qubit remains elusive due to challenges associated with obtaining the required nonlinearity. Consequently, there is a pressing need to mitigate the experimental prerequisites for generating an adequate GKP qubit for FTQC.

### 5.2 GKP states

GKP [412] qubits represent a significant development in the field of photonic quantum computing. These GKP qubits stand as a prominent contender for optical quantum computation due to two key factors: *firstly,* a significant subset of gates, operations, and measurements on GKP states can be executed with Gaussian resources. Gaussian resources are innately accessible and readily implementable on integrated photonic devices. *Secondly,* GKP qubits exhibit inherent resilience to noise and optical losses.

Gottesman, Kitaev, and Preskill introduced a method to encode a qubit using the position ($\hat{q}$) and momentum ($\hat{p}$) quadratures of an oscillator to correct errors resulting from small deviations in these quadratures [412]. The ideal GKP code states are represented by Dirac combs in the $\hat{q}$ and $\hat{p}$ quadratures, denoted as $|0\rangle_{\text{GKP}} = \sum_{k=-\infty}^{\infty} |2k\sqrt{\pi}\rangle_{\hat{q}}$ and $|1\rangle_{\text{GKP}} = \sum_{k=-\infty}^{\infty} |(2k+1)\sqrt{\pi}\rangle_{\hat{q}}$, respectively. However, these ideal states are non-normalizable and necessitate infinite squeezing [451].

In practice, physical GKP code states are approximated with finite squeezing. The basis of the GKP qubit with finite squeezing comprises Gaussian peaks of width $\sigma$ and separation $\sqrt{\pi}$ within a larger Gaussian envelope of width $1/\sigma$. The approximate code states $|\widetilde{0}\rangle$ and $|\widetilde{1}\rangle$ are defined as follows:

$$|\widetilde{0}\rangle \propto \sum_{\eta=-\infty}^{\infty} \int e^{-2\pi\sigma^2\eta^2} e^{\frac{-(\hat{q}-2\eta\sqrt{\pi})^2}{2\sigma^2}} |\hat{q}\rangle \, d\hat{q}, \tag{3}$$

$$|\widetilde{1}\rangle \propto \sum_{\eta=-\infty}^{\infty} \int e^{-\pi\sigma^2(2\eta+1)^2/2} e^{\frac{-(\hat{q}-(2\eta+1)\sqrt{\pi})^2}{2\sigma^2}} |\hat{q}\rangle \, d\hat{q}. \tag{4}$$

Here, the squeezing level $\mathcal{S}$ is defined by $\mathcal{S} = -10\log_{10}(2\sigma^2)$. In the case of finite squeezing, there exists a finite probability of misidentifying $|\widetilde{0}\rangle$ as $|\widetilde{1}\rangle$ and vice versa [412]. The probability $E(\sigma^2)$ of misidentifying the bit value is calculated as

$$E(\sigma^2) = 1 - \int_{-\sqrt{\pi}/2}^{\sqrt{\pi}/2} dt \, \frac{1}{\sqrt{2\pi\sigma^2}} e^{(-t^2/(2\sigma^2))}, \tag{5}$$

corresponding to bit- or phase-flip errors on the GKP qubit [451].

GKP qubits [405, 412, 438, 452, 453] exemplify a distinct category of bosonic qubits, affording them unique advantages that are not available to conventional two-level qubits. When GKP qubits are employed to implement an advanced error-correcting code, like the surface code, an additional layer of analog information derived from GKP error correction [414] serves to identify GKP qubits that are more likely to have incurred errors. By integrating this supplementary analog data into the decoder of the higher-level code, a substantial enhancement in the performance of the ensuing error-correcting code can be achieved [157, 414, 415, 417, 426, 428, 429, 435, 436, 454]. It is worth noting that while it has been demonstrated that conventional two-level qubits can also derive some benefit from analog information in specific scenarios, such as qubit readout [455, 456], they do not possess the capability to access analog information in broader contexts, such as during gate operations.

## 6 Building photonic quantum computers

The primary technological components essential for photonic QIP can be delineated into three key stages. Firstly, it is imperative to efficiently generate single-photon states, necessitating indistinguishability of correlated states and precise control over DoFs. Subsequently, appropriate platforms have to possess the capability to manipulate single- or multi-photon states for executing unitary transformations. Finally, photons should be effectively measured utilizing suitable detection systems.

Photonic devices play a pivotal role in the manipulation, detection, and generation of light, emerging as essential contributors to the rapid progress in light-based quantum computing. Within the intricate framework of these devices, key components such as single photon sources [457–459], beam splitters [460–462], and phase shifters [463–466]. Effective frequency shifting and beam splitting play a crucial role in various applications, spanning microwave photonics [467–470], atomic physics [471, 472], optical communication [473, 474], and photonic quantum computing [120–126]. Photodetectors (PDs) [477, 655, 656], including photodiodes [478, 479], metal-semiconductor–metal photodetector [480–482], phototransistors [483], and phototubes [484] are fundamental components. Quantum interferometers [485–488], such as MZI [489–491], Michelson interferometer (MI) [492], Fabry-Pérot interferometer (F-PI) [493–496], Sagnac interferometer (SAI) [497–499], common-path interferometers (C-PI) [500, 501], and fiber interferometers (FI) [502, 503]. Additionally, components like entanglement sources (ESs) [504, 505], QRs [506, 507], and waveguides [508–510] work in harmony, collectively establishing quantum photonic devices, that shape the transformative landscape of photonic quantum computing. Herein, we explore such crucial photonic device components.

### 6.1 Light squeezers

Squeezed light [511] generators constitute pivotal building blocks for QIP grounded in photonic technologies. The process of squeezing serves as a critical asset for quantum sensing applications [512] and a diverse range of quantum computing algorithms [119, 133, 513, 514]. Notably, significant efforts have been devoted to the development of scalable embodiments of these sources [47, 515–530].

Squeezing' of light [531–536] finds application in manipulating quantum noise distributions to enhance quantum sensing and various other utilization. Light squeezing constitutes a quantum effect enabling the reduction of noise variance in one observable of the electromagnetic field, such as its amplitude, below the quantum vacuum noise level. This reduction, however, comes at the cost of an increased noise variance in its conjugate observable, namely, its phase. Noteworthy instances include the application of optical laser light squeezing in gravitational wave detection [537] and microwave light squeezing for dark matter particle exploration [538]. Presently, research is concentrated on investigating and demonstrating diverse means of generating squeezed light resources. These resources can subsequently interact with a variety of optical cavities, opening avenues for their deployment in diverse applications [521, 539, 539–555].

### 6.2 Quantum light sources

A singular-photon emitter represents a facilitating technology in quantum simulation [99, 556], device-independent QCOMM [557], LOQC [73], and MBQC [558]. SPCs are integral components in quantum applications, constituting a unique class of quantum light sources that emit precisely one photon at a defined moment. These sources exhibit well-defined characteristics, including polarization and spatial-temporal mode [458, 459, 559].

Single photons are generated through various means such as single-photon emitters, parametric down-conversion (PDC), and QDs [560, 561]. It is crucial for these single photons to demonstrate uniform polarization, spatial-temporal mode, and a transform-limited spectral profile to ensure high visibility in quantum interference phenomena, exemplified by the Hong-Ou-Mandel-type interference [562]. This stringent criterion underscores the significance of coherent and controlled single-photon emission in advancing quantum technologies [563].

Spontaneous parametric down-conversion (SPDC) sources [564, 565] play a pivotal role in foundational quantum optics experiments, notably contributing to research recognized by the 2022 Nobel Prize in Physics for advancements related to entangled photons [109]. However, SPDC introduces intrinsic probabilistic elements and is unavoidably accompanied by multiphoton components. The single-photon efficiency of SPDC sources typically requires measures to suppress undesired two-photon emission. One strategy involves multiplexing multiple SPDC sources to enhance the efficiency of SPSs [566]. Alternatively, an innovative approach focuses on the direct generation of high-quality single photons from a two-level system. Among diverse platforms [567–574], semiconductor QDs [575] stand out as state-of-the-art SPSs, achieving an impressive overall efficiency of 57% [563]. This achievement is primarily attributed to a polarized microcavity developed by Wang et al. [576], featuring a polarization-dependent Purcell enhancement of single-photon emission that surpasses the 50% efficiency threshold. Future advancements, involving improved sample growth and enhanced collection efficiency, are anticipated to elevate single-photon efficiency beyond 70%, exceeding the requirements for universal quantum computing [577].

In the realm of quantum optics, another significant quantum light source is the squeezed state, denoting a quantum state where the uncertainty of the electric field strength for certain phases is smaller than that of a coherent state. Typically generated through intense pumping of nonlinear mediums [134], squeezed states, in conjunction with simple linear optical elements like B-Ss and PSs, facilitate the construction of CV quantum computing [134]. Gottesman, Kitaev, and Preskill [412] subsequently proposed a robust quantum error correction (QEC) code over CV to safeguard against diffusive errors. Presently, the record for squeezing is held by a 15 dB achievement from a type I optical parametric amplifier [578]. Various quantum experiments are underway, aiming towards the realization of large-scale CV quantum computing [393, 579, 580].

In recent decades, various approaches for SPSs have emerged [92, 458, 563, 575, 581–606, 608–629, 629–640]. Probabilistic sources encompass techniques such as PDC, which has been implemented in bulk crystals [612], microresonators [591], semiconductors [263, 636], and optical waveguides [92, 190, 581–590, 617–619]. Additionally, four-wave mixing (FWM) has been utilized in optical waveguides [592, 596], microdisks [594], and few-mode fibers [593, 637]. Deterministic sources [633], on the other hand, include architectures like trapped ions [613], color centers [611, 634], and QDs [185, 575, 624, 626, 638] employing materials such as InGaAs [195, 598–600, 606], GaAs [607], InAsP NW [609, 610], or InAs/GaAs structures [602–605].

## 6.3 Interferometers

An interferometer constitutes an optical device that exploits the phenomenon of interference, with a specific emphasis on optical interferometers designed for the manipulation of light [485–488]. Typically, the functionality of such a device is grounded in a specific operational sequence: initiating with an input beam, the beam undergoes division into two distinct beams using a B-S [463–466], often realized as a partially transmissive mirror. Subsequently, one or both of these beams may be subjected to external influences, such as alterations in length or refractive index within a transparent medium. Following this, the beams are recombined on another B-S, and the resulting beam's power or spatial characteristics can be utilized for diverse measurements [485–488].

The construction of interferometers often necessitates the use of high-quality optical components, with mirrors and optical flats, distinguished by a high degree of surface flatness [641–649], see Fig. 2. Various types of quantum interferometers [485–488] exist, encompassing the MZI [489–491, 641, 644], MI [492, 649], F-PI [493–496], SAI [497–499], C-PI [500, 501], and FI [502, 503, 643, 647].

## 6.4 Photodetectors

Photodetectors [475–477], serve the purpose of identifying and receiving light as well as various forms of radiation, spanning microwave and infrared wavelengths. In the realm of photonics devices, these instruments are commonly referred to as photon detectors due to their adeptness in discerning the stimulation of liberated charge carriers within these devices. Following the completion of the detection process, PDs typically generate an output, manifested as either an electric signal or a current [650].

Accurately detecting photons with high probability and reliability is a critical necessity for numerous applications, often posing a significant challenge to the overall efficiency of an apparatus. Given the extremely low energy of a single photon (approximately $10^{-19}$ J), PNR detectors necessitate high gain and minimal noise to effectively distinguish the correct photon count [39]. SPDs are designed to generate a measurable electrical signal upon stimulation by either one and only one incoming photon (referred to as PNR detectors) or by at least one photon.

The applications of PDs are extensive, encompassing utility in optical communication systems [654], optical radiometry, radiometry, photometry, spectrometers, interferometers, and optical sensors, among other domains [650]. PDs encompassing photodiodes [478, 479], metal–semiconductor-metal photodetectors [480–482], phototransistors [483], and phototubes [484].

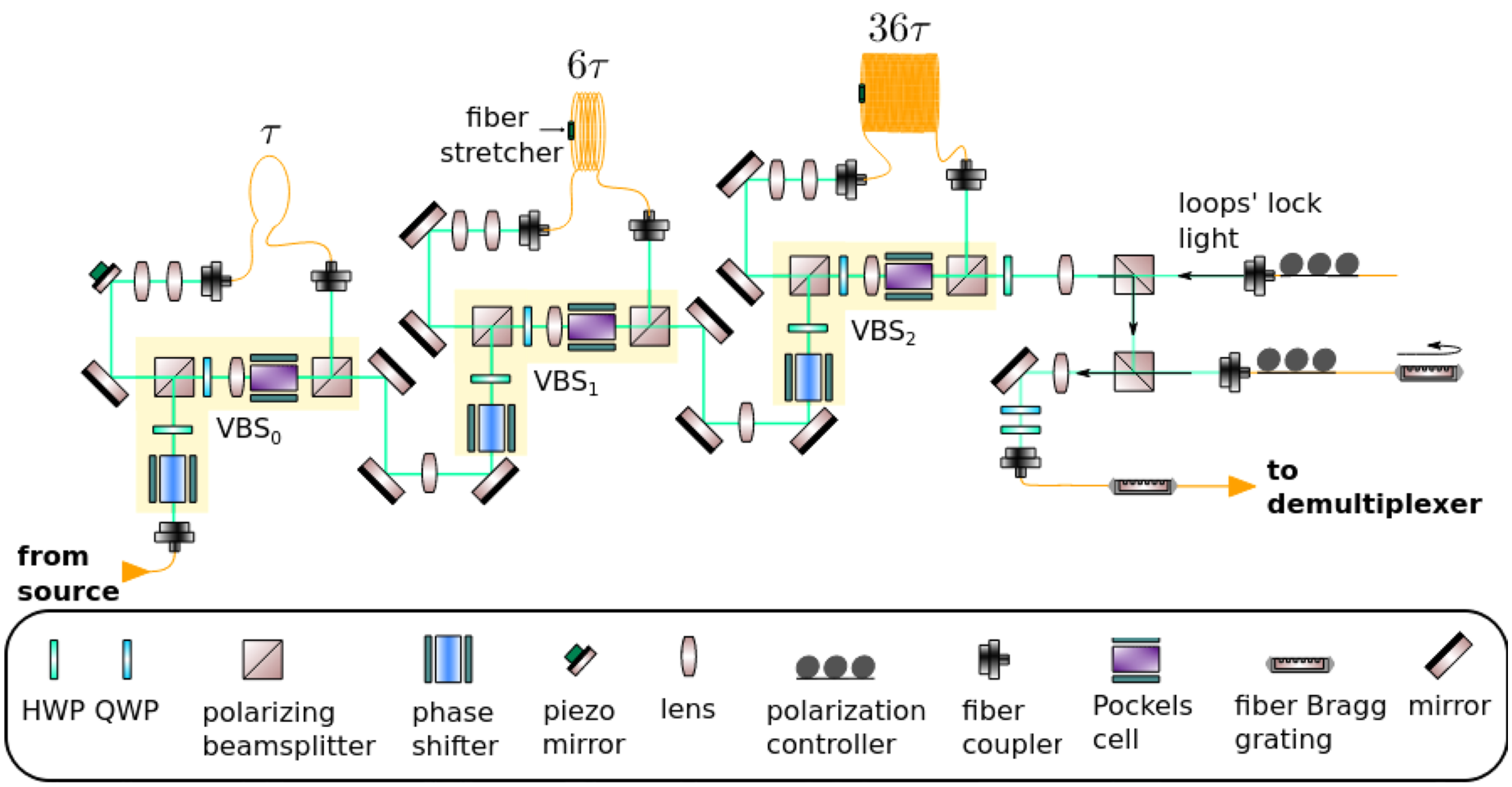

**Fig. 2** Programmable interferometer employing loop-based architecture. The variable beam splitter (VBS) is denoted within each loop, highlighted by the purple-shaded region. Regenerated under a Creative Commons License (http://creativecommons.org/licenses/by/4.0/) from [45]

Various types of detectors are employed for photon detection [655–663, 663–668], including both non-PNR and PNR detectors. Non-PNR detectors encompass technologies such as single-photon avalanche photodiodes (SPADs) made from materials like Ge-on-Si [669, 670], or InGaAs [671, 672], negative feedback avalanche diodes (NFAD) [673–675], QDs [676], superconducting nanowires [677–689], up-conversion detectors [690–695], and artificial Λ-type three-level systems [696]. PNR detectors include parallel superconducting nanowire single-photon detectors [697, 698], transition-edge sensors [79, 699–702], QD-coupled resonant tunneling diodes [676], multiplexed SPADs [703, 704], and organic field-effect transistors [705]. Efforts have also been made to integrate superconducting detectors into waveguide structures fabricated from various materials, such as Si [80, 706], GaAs [707–715], $LiNbO_3$ [701, 716], $Si_3Ni_4$ [717–723], and diamond [80, 701, 706–724]. For an in-depth understanding of the advancements in this field, see [664, 665, 725, 726].

## 6.5 Waveguides

A waveguide serves as a structural conduit that guides waves by channeling energy in a specific direction, minimizing losses during transmission [508–510]. Whether for electromagnetic waves or sound waves, waveguides efficiently transport these waves from one point to another in a medium with minimal energy dissipation. In the realm of optical waveguides, these structures exhibit spatial inhomogeneity to guide light, confining its propagation within a defined spatial region. Typically, a waveguide incorporates a region of heightened refractive index, known as cladding, compared to the surrounding medium [508–510].

Waveguides find diverse applications [727–739], such as facilitating optical fiber communications and playing a crucial role in photonic integrated circuits [73]. Moreover, these structures are instrumental in tasks like splitting and combining light beams, particularly evident in the functionality of integrated optical interferometers [73]. Diverse optical waveguide platforms have been devised for applications in integrated quantum photonics (IQP) [73, 74]. Theses encompass laser-induced silica waveguides [76, 169, 172, 740], silica-on-insulator ($SiO_2$) [57, 78, 118, 212], silicon-on-insulator (Si) [80, 81, 87, 741, 742], silicon nitride ($Si_3N_4$) [518, 721, 743, 744], gallium arsenide (GaAs) [745–747], lithium niobate (LN) [213, 701, 748], indium phosphide (InP) [91, 749], silicon oxynitride (hydex) [120], and various other substrates. Waveguide applications extend to medical diagnosis, health monitoring, light therapies [738] and diverse biomedical applications [739].

Figure 3 illustrates multiple instances of two-dimensional integrated waveguide meshes configurations, in which a unitary TBU (tunable basic unit) is spatially replicated to form distinct cells [64, 65, 70, 71, 651, 653, 750]. Additionally, configuration and emulation involving a hexagonal waveguide mesh for a triangular $4 \times 4$ universal interferometer are presented in Fig. 4. Where, Fig. 4a illustrates a $4 \times 4$ interferometer implemented through a triangular configuration of B-Ss, and Fig. 4b depicts a similar structure on a hexagonal waveguide mesh. Each B-S can determine a specific splitting ratio and a relative phase for the upper output. Algorithms have been developed for programming and configuring the triangular setup, allowing it to execute any desired linear unitary transformation [70]. GBS circuit for a photonic setup is shown in Fig. 5.

### 6.6 Linear optical networks

Linear optical networks (LOPNs) serve as foundational components in numerous protocols related to computation, communication, and sensing. LOPNs show versatility and adaptability in facilitating diverse functionalities across a spectrum of technological domains, such as neuromorphic and reservoir computing [751–754], optical simulation [755–757], and optical neural networks [60, 758].

LOPNs play a pivotal role in QIP, with the interferometer serving as a unitary transformation on either the single-photon Fock state or the single-mode squeezed state. A seminal work by Reck et al. [70] established that a universal unitary transformation could be achieved through a triangular arrangement of B-Ss and PSs. The optical depth in this configuration is $2(m-1)-1$, and the number of B-Ss is determined by $m(m-1)/2$, where 'm' indicates the number of modes. Clements et al. [71] explored the equivalence of an interferometer with a rectangular configuration to its triangular counterpart, resulting in a reduction of optical depth to $m-1$ and a decrease in the number of B-Ss to $(m(m-2)+2)/2$. This configuration proves to be more compact and robust due to its symmetrical design.

In the context of BS, an effective LOPNs necessitates the simultaneous integration of high transmission, Haar randomness, and significant spatial and temporal overlap. Various implementation strategies include time-bin loops [45, 237], micro-optics [43, 44, 95, 99], and integrated photonic circuits [55, 56, 82]. Time-bin loops and integrated on-chip circuits offer programmability but are challenged by substantial losses. On the other hand, micro-optics [43, 44, 95, 99], while exhibiting superior transmission efficiency, lacks programmability. Addressing the conundrum of reducing losses while maintaining programmability remains a significant and ongoing objective for future advancements in LOPNs [759–762].

## 7 Towards scalable photonic quantum computers

Photonic qubits and optical modes constitute the cornerstone of optical quantum computing, utilizing DV and CV technologies, respectively. DV systems encode quantum information in discrete states of photons, such as polarization or path, which are manipulated as qubits. On the other hand, CV systems utilize the continuous parameters of the electromagnetic field, such as amplitude and phase, to encode information in optical modes.

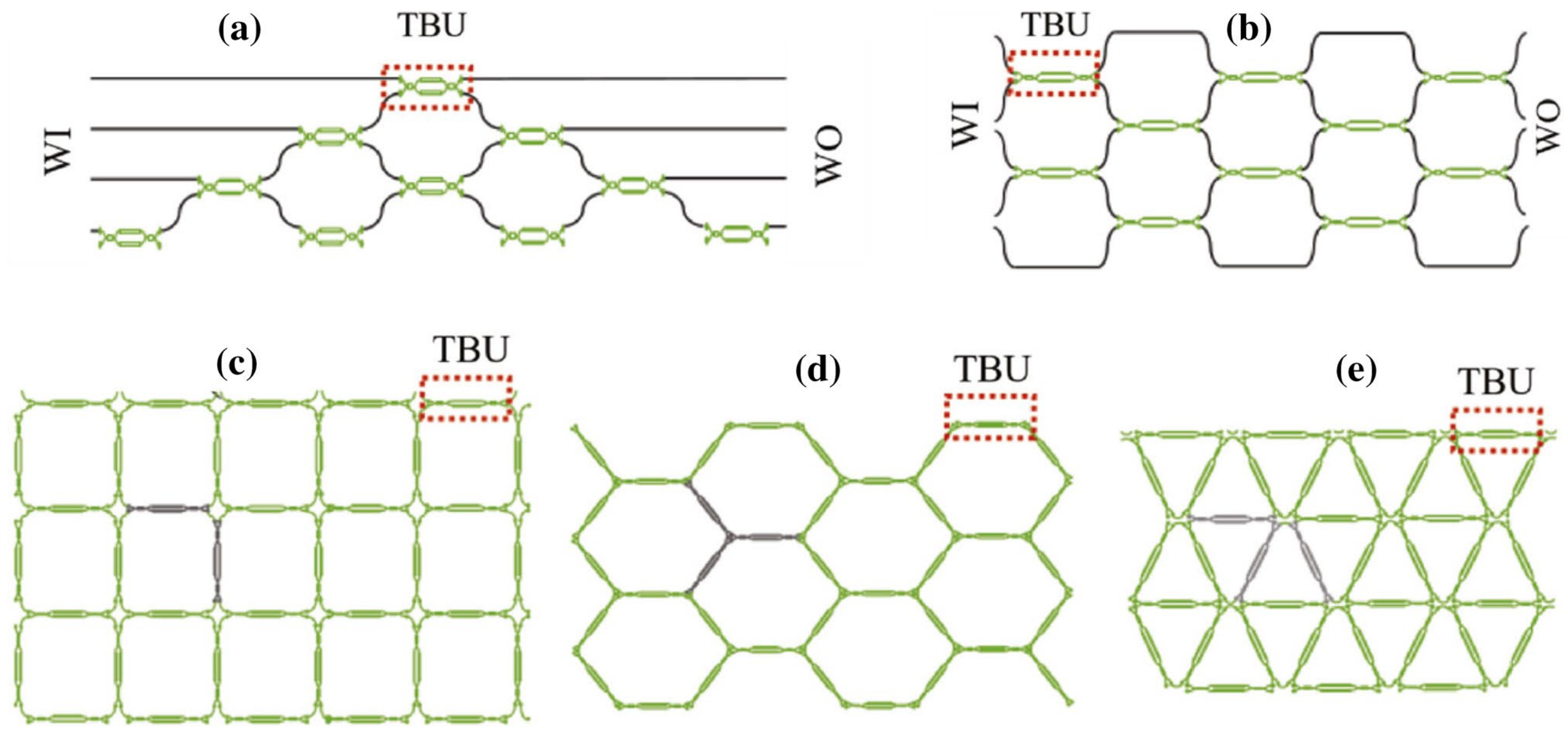

**Fig. 3** Various configurations of beam splitters employed for the realization of integrated waveguide meshes. **a** Triangular feedforward initially proposed by Reck et al. [70] and subsequently reformulated [651], **b** rectangular feedforward proposed by Clements et al. [71], **c** squared feedforward/backward [64], **d** hexagonal feedforward/backward [652, 653], and **e** triangular feedforward/backward [652, 653]. WI/WO signifies waveguide inputs/outputs. Reproduced under a Creative Commons license (https://creativecommons.org/licenses/by-nc-nd/4.0/) from [652]

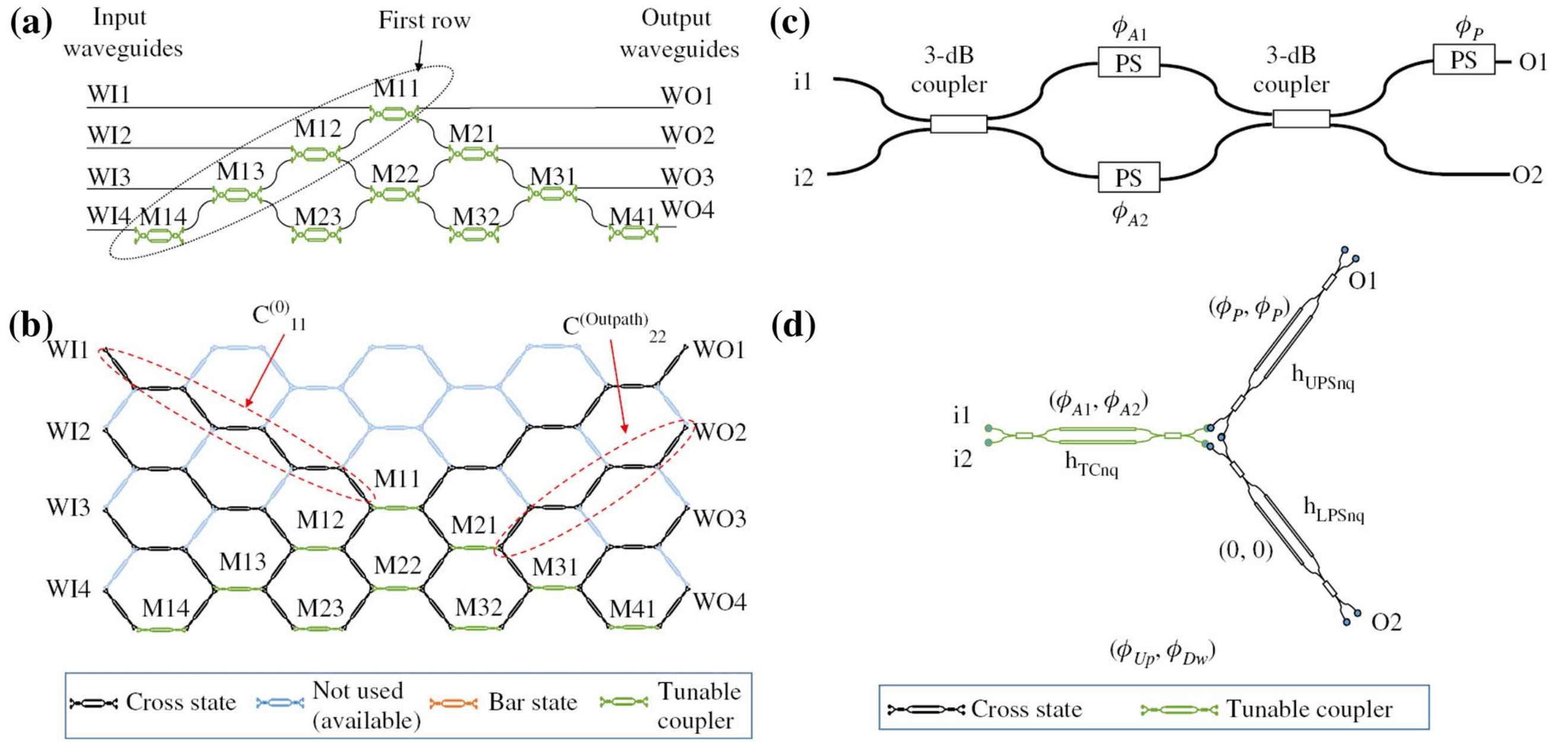

**Fig. 4** Design and simulation involving a hexagonal waveguide mesh for a triangular $4 \times 4$ universal interferometer. **a** Classical triangular configuration and **b** implementation of a $4 \times 4$ interferometer using a hexagonal mesh. **c** Beam splitter for the classical approach and **d** the corresponding beamsplitter implementation utilizing three TBUs for the hexagonal waveguide mesh. Reproduced under creative commons license (https://creativecommons.org/licenses/by-nc-nd/4.0/deed.en) from [652]

## 7.1 Photonic-qubit (DV) approach

The DV approach in photonic quantum computing utilizes distinct quantum states of photons to encode and process quantum information within discrete levels ($\sum_{j=0}^{n-1} \gamma_j |j\rangle$) defined in Hilbert spaces [25]. This framework has proven useful in the implementation of various quantum computing and cryptography protocols, leveraging photonic qubits ($n = 2$) and qudits ($n > 2$). Photons possess multiple DoFs that are particularly suited for these tasks, allowing encoding information through specific photon DoFs. Alternatively, polarization serves as a natural 2-level system equivalent to a qubit's Hilbert space. Moreover, photons can be manipulated to occupy high-dimensional DoFs, enabling the encoding of qudits [39, 763, 764]. These higher-dimensional DOFs include DVs such as different time intervals or frequencies [249, 765, 766], optical paths [73, 97, 104], spatial modes supported by optical fibers [767, 768], as well as orbital and transverse momentum [357, 769]. Such versatility in photon manipulation allows for robust implementations of quantum algorithms and secure communication protocols.

## 7.2 CV approach

Unlike discrete qubits or qudits, CV quantum optics exploits physical observables that assume continuous values within the phase space of quantum harmonic oscillators associated with electromagnetic field modes [134, 770]. This framework enables the encoding and manipulation of quantum information using continuous properties such as amplitude and phase of photons, rather than discrete levels. CV quantum computing has been pivotal in proposing protocols for quantum computation, cryptography, and recently, algorithms for sampling based on photon counting experiments [133, 412, 556, 771–774].

In CV optical quantum computation [134], the fundamental unit is not the photonic qubit but the optical mode, also known as qumode. Information is encoded using the continuous values of optical modes, referred to as quadratures. The CV computational model was originally proposed in [134]. CV quantum computation relies on manipulating the distributions of quadrature values, categorized into linear (Gaussian) and nonlinear (non-Gaussian) transformations. Linear transformations, being Gaussian, can be implemented deterministically and have been successfully demonstrated. In contrast, direct realization of non-Gaussian transformations proves non-trivial due to the requirement for exceptionally high nonlinearity, a characteristic not readily available in conventional optical media [776, 777].

However, by employing CV quantum teleportation techniques, even non-Gaussian transformations can be deterministically realized. This determinism stands as a noteworthy advantage of the CV system. Notably, CV quantum teleportation, as demonstrated by Furusawa et al. in [579]

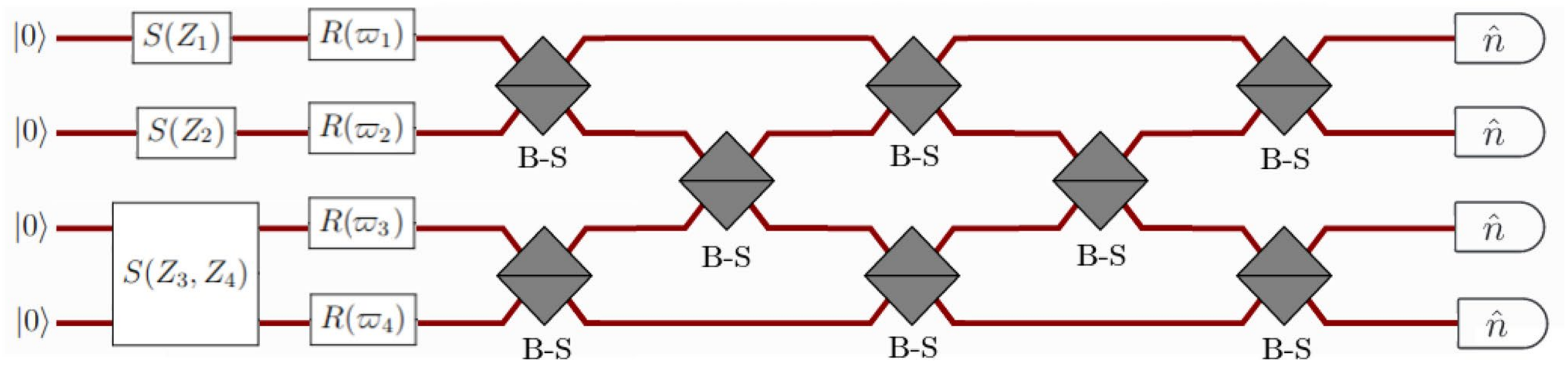

**Fig. 5** A photonic configuration employing a GBS circuit. The quantum modes are initialized in Gaussian states derived from the vacuum through squeezing operations denoted as $S(Z_i)$. Subsequently, an interferometer is introduced, comprising phase shifters denoted as $R(\varpi) = e^{i\varpi_l}$ and beam-splitters (B-Ss). Finally, photon number-resolving measurements are conducted in each mode [775]

and extensively researched since then, consistently achieves success in a deterministic manner. Consequently, CV optical quantum computation can leverage the inherent advantages of optical systems without relying on probabilistic operations.

While quantum teleportation might seem initially confined to transmitting input states, modifications to quantum entanglement and Bell measurements give rise to MBQC [16, 778]. CV optical quantum teleportation, in contrast to many physical systems, offers several advantages. Quantum entanglement can be easily and deterministically generated within the CV optical system, while CV Bell measurements can be executed with notable efficiency. Therefore, instead of adhering to the conventional gate model/circuit model utilized by matter-based qubits, it is more intuitive to contemplate CV quantum computation through the lens of quantum teleportation.

Presently, architectures for scalable and universal photonic quantum computing exhibit two contrasting paradigms. The first paradigm harnesses the remarkable scalability of CV entangled resource states to execute computations on DV information, particularly qubits, encoded within bosonic modes [413, 779]. While the production of CV resources for this approach can be deterministic and scalable, it necessitates the on-demand and deterministic generation of DV resources, imposing demanding hardware requisites that may be infeasible. The second paradigm centers on the generation of entangled resource states solely comprised of bosonic qubits [18, 415, 426, 780, 781]. Notable advancements have been made in the design and deterministic generation of CV cluster states in one [258, 259, 392], two [393, 394, 782–784], and higher dimensions [785–787]. While, other architectural schemes designed for various qubit encoding, such as the cat-basis encoding [780, 788], the GKP encoding [415, 451], and the dual-rail encoding [789], grapple with the non-deterministic generation of individual qubit states, particularly in the case of the former two, where the states possess intricate structures. Recently, another approach for fault-tolerant measurement-based photonic quantum computation exploits hybrid resource states, that possesses the advantage of CV-based schemes and yet is compatible with probabilistic GKP qubit sources [412], is proposed in [157].

## 8 Programmable photonic circuits

Photonic integrated circuits (PICs) [790–799] have emerged as a well-established and robust technological paradigm with diverse applications [800, 801]. Analogous to electronic integrated circuits, PICs are fabricated on chip surfaces; however, they modulate light rather than electrical signals, employing on-chip optical components like waveguides, beam couplers [802, 803], electro-optic modulators, photodetectors, and lasers [804]. While electronic circuits excel in digital computations, photonic circuits demonstrate proficiency in the transportation and processing of analog information [805]. Consequently, PICs find prevalent use in contemporary fiber-optic communications, while also proving instrumental in applications where light plays a pivotal role, such as chemical, biological, or spectroscopic sensing, metrology, as well as classical and QIP [790–801, 804, 806–811].

Programmable PICs operate on the principle of dynamically manipulating light flow within the chip during runtime, facilitated, for instance, by electrically controlled tunable beam couplers connected through optical waveguides [806]. This dynamic manipulation enables the distribution and spatial rerouting of light under software control [50, 651, 807]. These chips exhibit the capacity to perform various linear functions by interfering signals along distinct paths and can instantiate programmable wavelength filters, essential components for communication and sensor applications, as well as the manipulation of microwave signals in the optical domain [64, 468]. As these interconnected waveguide meshes scale up, they enable the execution of linear optical computations, including real-time matrix–vector products [60, 651, 808]. Such computations are pivotal in QIP [46, 71, 809, 810], neuromorphic computing, and artificial intelligence [60, 808]. Notably, rapid advancements in programmable PICs technologies for these applications are already underway. Similar to their electronic counterparts, the programmability of these circuits allows for the re-configuration of functionality at runtime, thereby reducing economic and technological barriers and providing a pathway to upgradability [806].

Programmable photonics constitutes a wideband analog technology that empowers the programming of signal processing tasks, leveraging the capacity of photonic circuits to manipulate multiple optical interferences. Fundamentally, this involves independently configuring the amplitude and phase characteristics of interfering signals. For the former, tunable couplers or MZI are employed, while the latter involves the use of a phase shifter. These serve as fundamental building blocks, subject to programming through external electronic signals [804, 811].

The amalgamation and interconnection of these foundational components enable the realization of programmable PICs with varying degrees of complexity and functionality, categorizable into three distinct hardware families, see Fig. 6. The most rudimentary configuration for programmable PICs, known as the reconfigurable application-specific photonic integrated circuits (ASPICs), retains essential features of fixed designs while introducing a measure of reconfigurability, allowing for the programming of operational and bias points governing circuit response without altering the overall chip functionality. A second hardware family comprises multipoint interferometers, structured on 2D fixed topologies constituted by tunable interferometers that can be programmed to simulate any linear feedforward arbitrary unitary matrix transformation. Finally, photonic waveguide meshes, founded on open 2D topologies adhering to regular geometric patterns, possess the capability to emulate any reconfigurable ASPIC and multiport interferometer, while additionally facilitating the implementation of both feedforward and feedbackward transformations [804, 811].

In programmable PICs, the regulation of light flow is achieved through the interconnection of waveguides forming a mesh structure, employing $2 \times 2$ blocks referred to as 'analogue gates'. These gates serve as on-chip analogs to free-space optical B-Ss. The configurability and potential functions of the programmable circuit are dictated by the mesh connectivity, determined by the arrangement of these analogue gates. Certain architectures permit the execution of arbitrary matrix operations [47, 50, 52, 57, 60, 64, 71, 97, 127, 468, 642, 651, 802, 807, 814–823] and possess the capability to autonomously adapt to evolving problem scenarios [50, 127, 651, 802, 803, 807, 823].

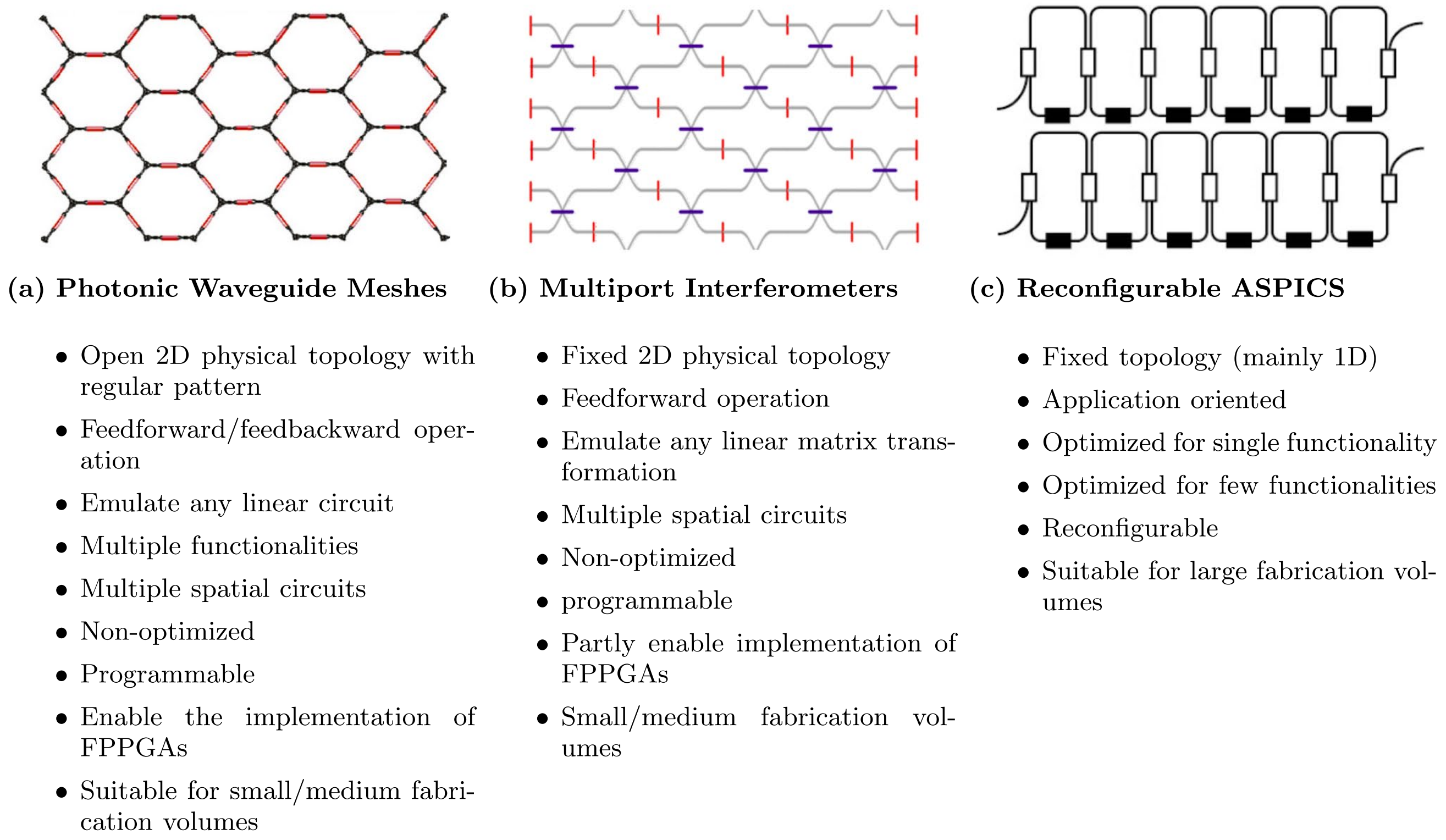

**Fig. 6** Programmable integrated photonics hardware categorization. **a** Photonic waveguide meshes. **b** Multipoint interferometers, and **c** the reconfigurable ASPIC. Underscores its capacity to facilitate analog signal processing tasks, leveraging the inherent capability of photonic circuits to adeptly handle numerous instances of optical interference. The photonic waveguide meshes can emulate either multipoint interferometers or ASPIC. While multipoint interferometers can emulate some ASPIC. Adapted from [811]

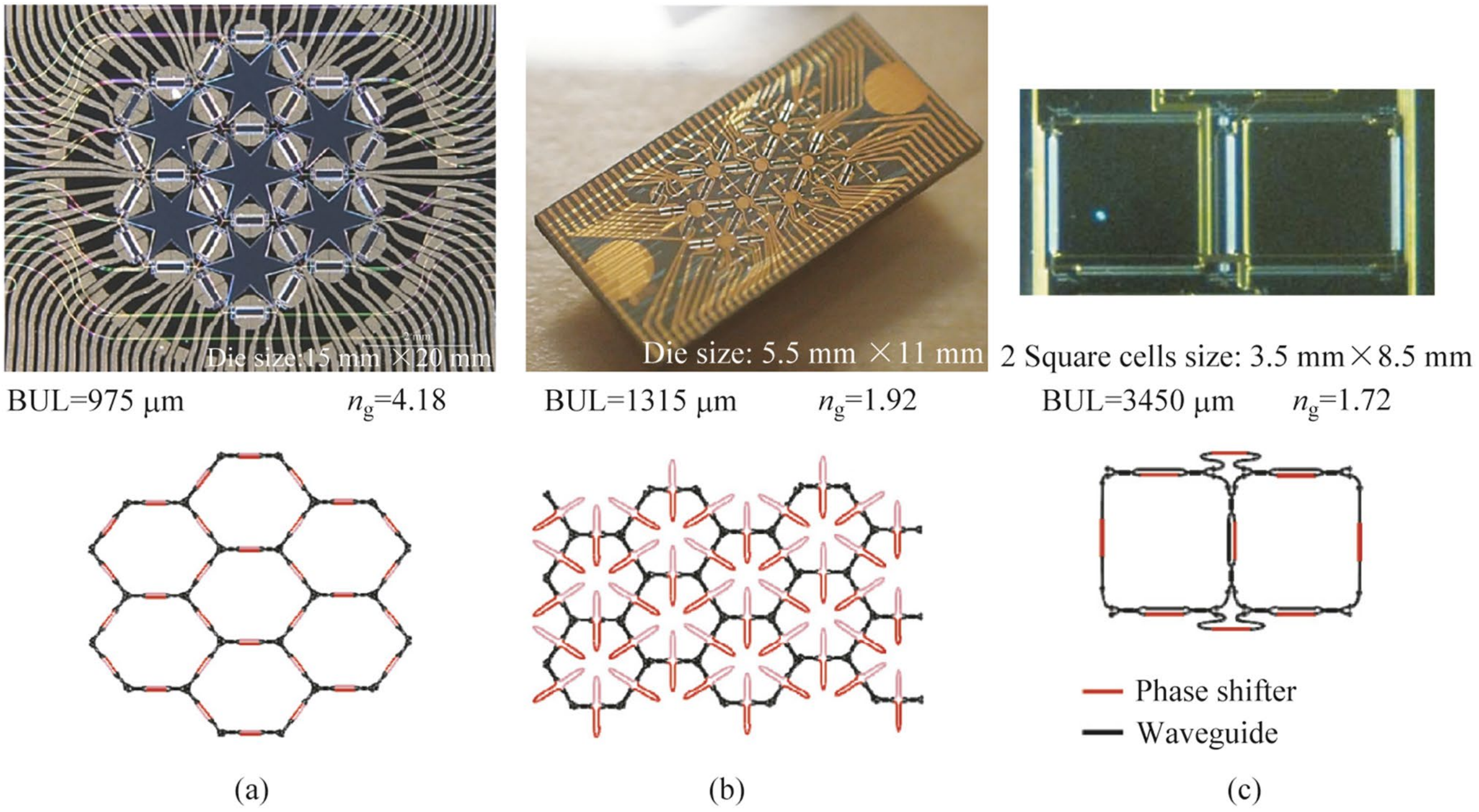

**Fig. 7** Representation of chip images and fabricated layouts for various feedforward and backward waveguide meshes employing diverse material platforms and cell geometries. **a** Hexagonal configuration implemented in silicon [65]. The chip consisting of seven hexagonal cells (equivalent to 30 thermally tuned TBUs) produced in Silicon on Insulator. The device was manufactured at the Southampton Nanofabrication Centre, University of Southampton. Silicon on insulator (SOI) wafers, featuring a 220-nanometer-thick silicon overlayer and a 3-micrometer-thick buried oxide layer, (for more comprehensive information on fabrication and testing, refer to [65]). **b** hexagonal configuration in $Si_3N_4$ incorporating a modified TBU scheme [812], and **c** square configuration in $Si_3N_4$ [64]. Regenerated under creative commons license (https://creativecommons.org/licenses/by-nc-nd/4.0/) from [652]

The classification of waveguide meshes can be broadly categorized into two classes: (i) forward-only, characterized by unidirectional light flow from one side of the mesh to the other [50, 70, 71, 651, 803, 808], and (ii) recirculating, wherein light can be routed in loops, including back to the input ports [64, 642, 806, 820]. Notably, both architectures share common building blocks, consisting of waveguides, 2 × 2 couplers, and optical phase shifters, collectively forming the foundation for the analogue optical gates in these circuits [804]. Distinct functionalities are subsequently attained by choosing the appropriate trajectory within the mesh. The essential regular and periodic geometries are achieved through the replication of square [64], triangular [653], or hexagonal [65, 642, 653] unit cells in the formation of 2D integrated waveguide meshes [64, 65, 70, 71, 651, 653, 750]. Figure 7 presents a photographic representation of the chip and the constructed layout for diverse feedforward/backward waveguide meshes employing distinct material platforms and cell geometries.

Programmable photonics has proven effective across a range of applications, such as functioning as signal accelerators for machine learning hardware, quantum processing, and supporting general-purpose optical signal processing [804, 811]. The advent of programmable PICs stands to revolutionize the utilization of coherent light for information manipulation. Upon widespread industrial implementation, programmable PICs hold the potential to significantly reduce the lead time for photonic chip production from months to days [804], thereby mitigating considerable non-recurrent engineering costs. This shift also signifies a transition in product development emphasis from hardware to software [824]. Figure 8 highlights significant achievements in quantum integrated photonic over the past decade (2012–2022) [813]. For more understanding of programmable PICs, readers are directed to [804, 806, 811, 813, 825].

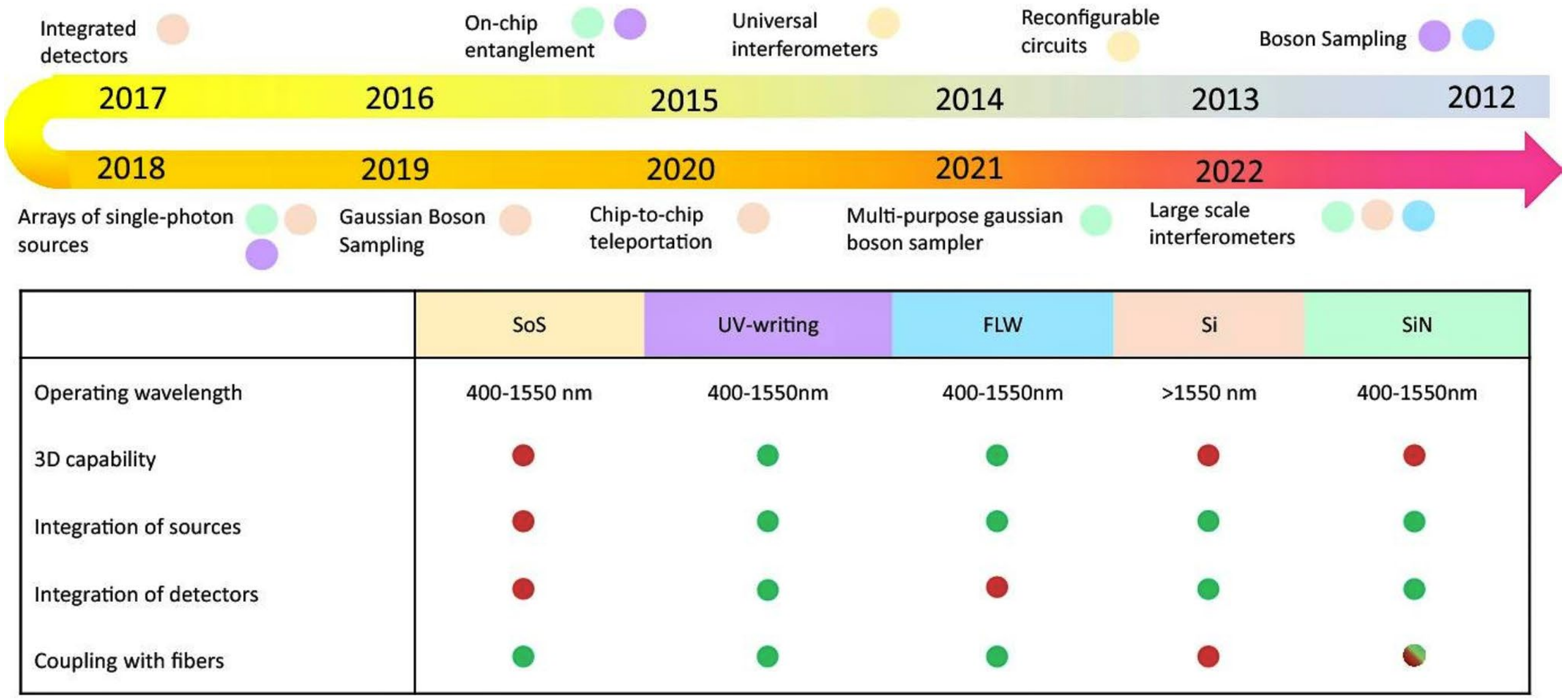

**Fig. 8** Integrated photonics in quantum technologies: A review of key fabrication technologies focusing on circuit design, operating wavelengths, integration of sources and detectors, and connection with external optical fibers. Each method is represented by a different color on the timeline. The timeline highlights significant achievements in quantum integrated photonics (QIP) over the past decade (2012–2022). Colored circles indicate the manufacturing technologies used: silica-on-silicon (SoS), silicon nitride (SiN), femtosecond laser writing (FLW), silicon (Si), and UV writing. Reproduced from [813] under a Creative Commons Attribution 4.0 International License (http://creativecommons.org/licenses/by/4.0/)

## 9 Photonic quantum communication and internet

Harnessing photons as "flying qubits" to manipulate quantum information becomes imperative for tasks in communication-centric QIP, such as linking quantum computers and enabling distributed processing. The compelling advantages of utilizing photons as carriers of information are manifest (refer to Sect. 4), owing to their inherent purity and resilience against decoherence-attributes that position photons as exemplary quantum systems [826].

### 9.1 Reconfigurable photonic technology for quantum communication

Reconfigurable quantum photonic components play a pivotal role in the manipulation of quantum states of light, a fundamental requirement for QIP in QCOMM and Internet. This manipulation is effectively achieved through the utilization of readily available passive and active integrated photonics components [827]. In the realm of practical QCOMM systems, the necessity for single-photon sources and entangled photon sources is not absolute [827]. The decoy-state protocol [828–830], posits that weak coherent pulses can serve as a viable substitute for single-photon states in the majority of prepare-and-measure QKD applications. Consequently, the generation of integrated photon sources can be straightforwardly accomplished by attenuating coherent pulses generated by on-chip lasers. The realization of such photon sources has been successfully demonstrated in various chip-based QKD systems [91, 831, 832].

In the context of a typical QCOMM system, photons are managed across various DoFs, including polarization, phase, spatial, spectral, and temporal domains. Consequently, the construction of versatile building blocks capable of influencing these photon DoFs becomes imperative [827]. These building blocks encompass polarization splitters/rotators (see Fig. 1 in [833]), PSs (see Fig. 1 in [834]), intensity modulators (see Fig. 2 in [835]), directional couplers (see Fig. 1 in [836]), multi-mode interferometers (MMI) [837] (as depicted in Fig. 9 (a,b)), ring resonators (see Fig. 9c) [838], and delay lines (see Fig. 1 in [839]).

Notably, PSs find realization through the thermo-optic effect for low-speed applications [47, 834] and the Pockels electro-optic effect for high-speed applications [835, 840]. Demonstrations of such devices span various integrated platforms, such as an ultraviolet-written silica-on-silicon (UWSOS) photonic chip for quantum teleportation featuring thermo-optic PSs [841], a GaAs quantum photonic circuit with a tunable MZI depending on the Pockels effect [746], a re-programmable linear optical circuit incorporating an array of 30 silica-on-silicon waveguide directional couplers with 30 thermo-optic PSs (depicted in Fig. 1 in [46]), and a large-scale silicon photonics quantum circuit integrating 16 SFWM photon-pair sources (see Fig. 9d), 93 thermo-optical PSs, and 122 MMI B-Ss [97]. Utilizing on-chip modulators, grounded in the quantum-confined Stark effect [831] or the free-carrier dispersion effect [842, 843], presents a viable approach for pulse generation and qubit encoding with the potential to attain frequencies reaching the gigahertz range. In the context of polarization-encoding protocols, specifically for the generation of BB84 polarization states, modulators employing polarization rotators and polarization B-Ss have been developed and validated [184, 843, 844].

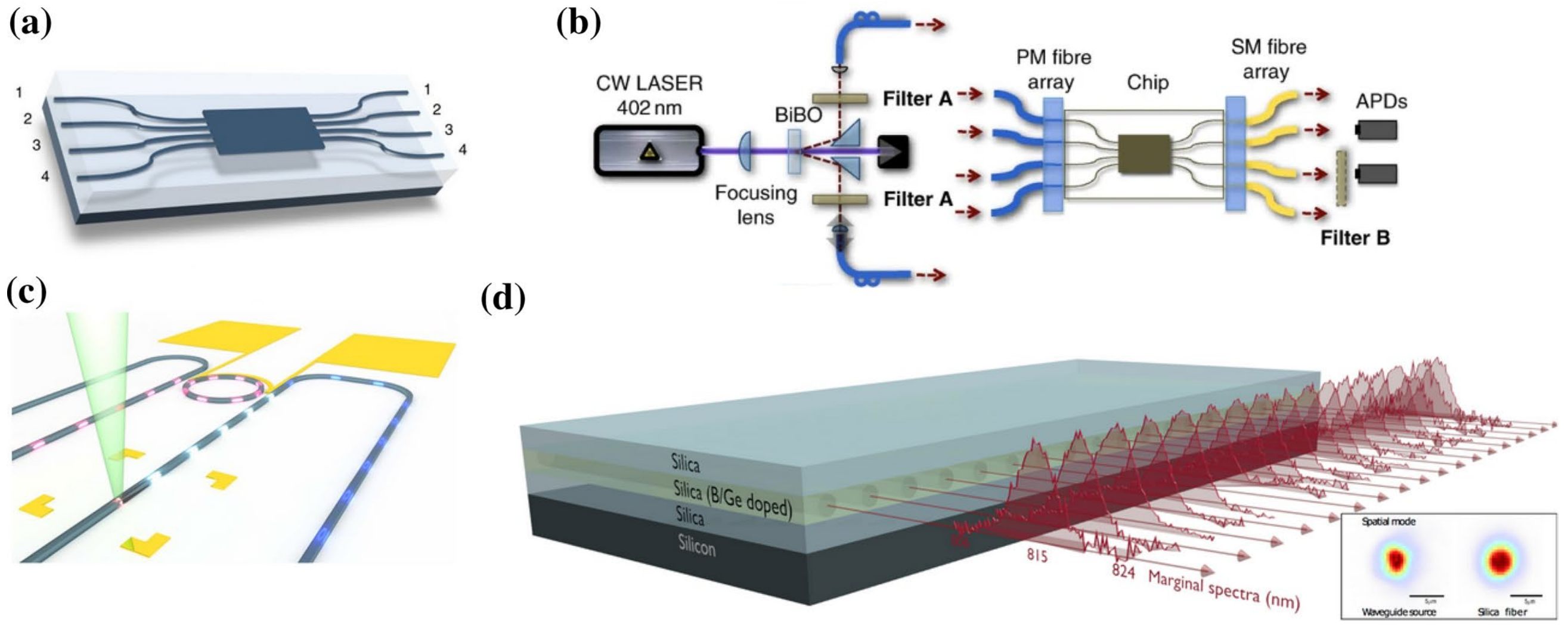

**Fig. 9** **a** Schematic representation of a $4 \times 4$ multi-mode interferometers integrated chip. **b** Experimental arrangements for conducting quantum interference measurements involving two photons in MMI devices. The PDC source utilized in these experiments incorporates two instances of Filter A, featuring FWHM of 2 nm for the $2 \times 2$ MMI measurements and 0.5 nm FWHM for the $4 \times 4$ MMI measurements. The inclusion of Filter A is essential to ensure the indistinguishability of single photons. Additionally, in the $2 \times 2$ MMI measurements, Filter B with a 0.5 nm FWHM was introduced to enhance the coherence length of the photons. The experimental setup includes various components such as continuous wave (CW) lasers, bismuth borate (BiBO) crystals, polarization-maintaining (PM) fibers, single-mode (SM) fibers, and silicon single-photon avalanche photodiodes (APDs). **c** Schematic representation of a fabricated hybrid quantum photonic circuit, seamlessly integrated with an on-chip tunable ring resonator filter. **d** An array of heralded single-photon sources (HSPSs) based on SFWM. This array is realized through the fabrication of straight waveguides using UV-laser writing on a germanium-doped silica-on-silicon photonic chip. Each individual straight waveguide within this series constitutes its own HSPS. **a**, **b** Reproduced from [837] under a Creative Commons license (http://creativecommons.org/licenses/by-nc-nd/3.0/). **c** Reproduced from [838] under a Creative Commons license (https://creativecommons.org/licenses/by/4.0/). **d** Reproduced from [92] under a Creative Commons license (https://creativecommons.org/licenses/by/4.0/)

In addition to the aforementioned elements, the integration of further components is essential for optical connectivity between quantum photonic chips and optical fibers. One-dimensional grating couplers and off-plane coupling prove effective when dealing with a single input or output polarization [845]. Alternatively, in scenarios involving multiple polarizations and a broader spectral range, edge couplers such as inverted tapers for butt coupling can be adopted [846]. Furthermore, two-dimensional grating couplers that support multi-polarization operation have been demonstrated to convert path-encoded qubits into polarization-encoded qubits, a format more suited for propagation in optical fibers [90, 187].

Figure 10 presents different chip-based QCOMM systems have been developed to support advanced QKD protocols. A depiction of an experiment involving the distribution of entanglement in a high-dimensional space is presented in Fig. 11. This experimental setup, involved utilizing an exceptionally intense source of hyperentangled photons, with a detection apparatus (referred to as *Alice*) stationed at the Institute for Quantum Optics and Quantum Information (IQOQI), and a receiving station (referred to as *Bob*) situated at the University of Natural Resources and Life Sciences (BOKU) in Vienna [202]. An experimental setup of light-based integrated quantum random number generators (QRNG) is shown in Fig. 12.

## 9.2 Secure quantum communication systems

Quantum communication employs the principles elucidated in quantum mechanics [847] to facilitate the transmission of quantum information, thereby facilitating significant enhancements in security protocols, computational capabilities, sensing technologies, and metrological techniques [827, 848]. This domain encompasses a wide array of technologies and practical applications, spanning from advanced laboratory investigations to tangible commercial implementations. Among the most prominent examples is QKD [160, 849, 850], wherein the fundamental concept involves utilizing the quantum properties of photons to establish confidential keys between geographically distant entities. The inherent property of quantum non-cloning theorem confers upon the communicating parties the capacity to identify any illicit attempts by third-party eavesdroppers to intercept or decipher the transmitted key [281, 851]. By grounding its security measures in the fundamental tenets of quantum physics rather than relying solely on computational complexity, QKD emerges as a highly sought-after solution to mitigate the escalating risks posed by the advent of quantum computing hardware and algorithms [827].

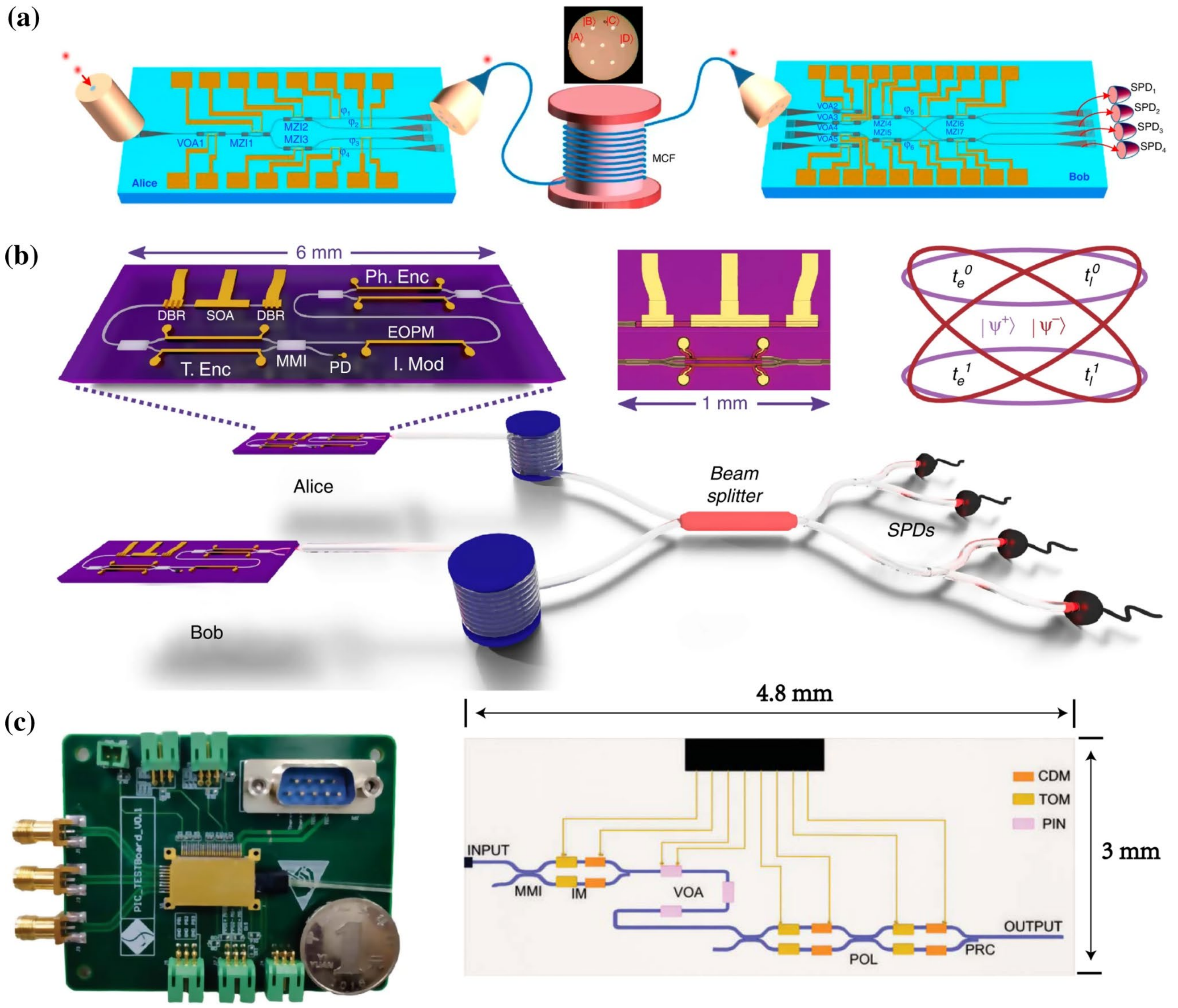

**Fig. 10** Various chip-based QCOMM systems that have been developed to support advanced QKD protocols. **a** Silicon-photonic-integrated circuit designed for noise-tolerant high-dimensional QKD [852]. **b** InP transmitter chips utilized for generating time-bin encoded BB84 weak coherent states for MDI-QKD [831]. **c** A packaged silicon photonic MDI-QKD transmitter chip connected to a compact control board and its corresponding schematic [843]. Reproduced from: (a) [852], under a Creative Commons license (https://creativecommons.org/licenses/by/4.0/); (b) [831], under a Creative Commons license (https://creativecommons.org/licenses/by/4.0/); (c) [843], under a Creative Commons license (https://creativecommons.org/licenses/by/4.0/)

Despite the contentious discussions surrounding its practical security implications [853], QKD is spearheading the transition towards tangible real-world implementations [155]. Notably, successful demonstrations of QKD experiments have been conducted using fiber-based and satellite-to-ground setups, achieving distances of over 800 km in ultra-low-loss optical fiber [160] and 2000 km in free space [103], respectively. Remarkably, advancements in technology have elevated the maximal secure key rate for a single channel to surpass 110 Mbit/s [161]. Additionally, several operational QKD networks have been established across various regions including Europe [854–856], Japan [857], China [858, 859], the UK [860], and others. Moreover, extensive efforts have been dedicated to scrutinizing the security of practical QKD systems, aiming to surmount existing technical constraints [154, 155, 853]. Collaborative endeavors have also integrated post-quantum cryptography with QKD protocols to ensure both short-term authentication security and long-term key security [861].

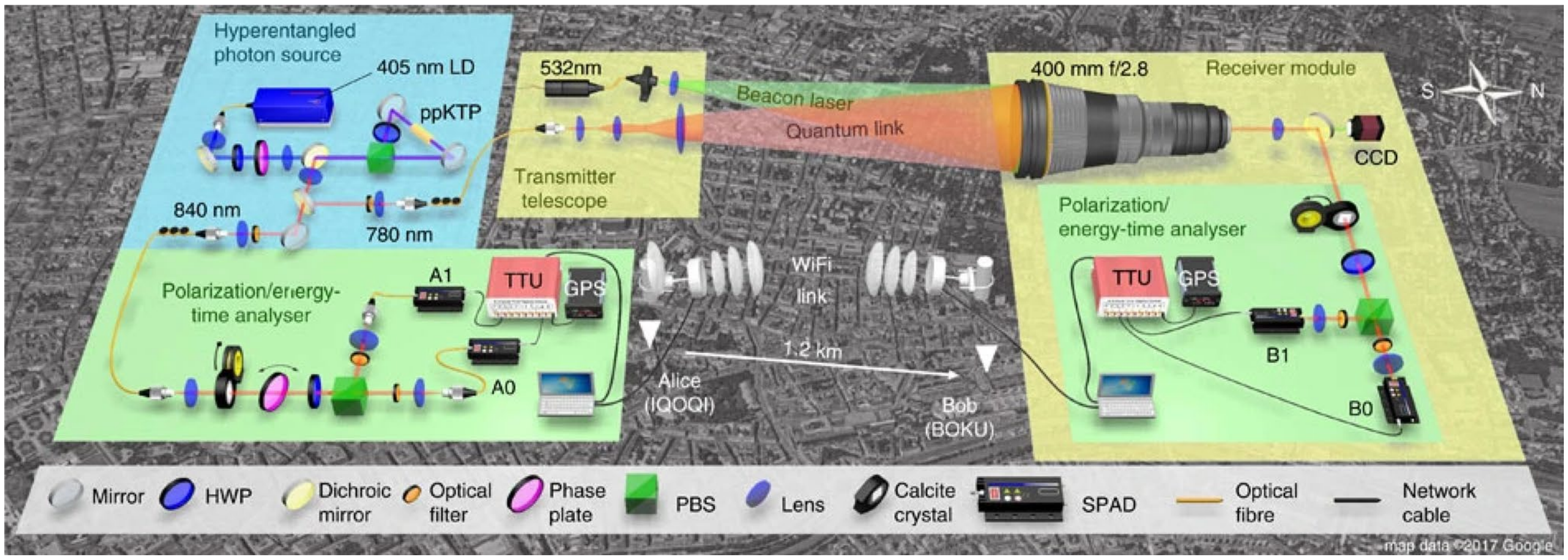

**Fig. 11** Distribution of entanglement in a high-dimensional space. A hyper-entangled photon source, situated within the confines of the Institute for IQOQI Vienna laboratory, employed SPDC within a periodically poled $KTiOPO_4$ (ppKTP) crystal positioned at the focal point of a Sagnac interferometer. This source was driven by a continuous-wave 405-nm laser diode (LD). Photon pairs, entangled in both polarization and energy-time domains, exhibited central wavelengths of approximately $\lambda_B \approx 780$ nm and $\lambda_B \approx 840$ nm, respectively. Photon $A$ was directed to *Alice* at IQOQI Vienna via a short fiber optic link, while photon $B$ was transmitted to *Bob* at the BOKU in Vienna, through a 1.2-km-long free-space link utilizing a transmitter telescope atop the institute's roof. *Bob*'s photon reception utilized a large-aperture telephoto objective with a focal length of 400 mm. A 532-nm beacon laser was isolated from the hyperentangled photons using a dichroic mirror and focused onto a CCD image sensor to facilitate link alignment and atmospheric turbulence monitoring. Analysis of polarization or energy-time bases was conducted through *Alice*'s and *Bob*'s analyzer modules, employing a half-wave plate (HWP) and a polarizing beam splitter (PBS) with a single-photon avalanche diode (SPAD) in each output port. A birefringent crystal tilt introduced an additional phase shift in *Alice*'s module. Optional calcite crystals before the PBS enabled the introduction of polarization-dependent delay for Franson interference measurements in the energy-time domain. Single-photon detection events were recorded using a GPS-disciplined time tagging unit (TTU) and stored locally for subsequent processing. *Bob*'s measurement data were wirelessly transmitted to *Alice* in real-time via a classical WiFi link for the identification of photon pairs. Reproduced from [202] under a Creative Commons Attribution 4.0 International License (http://creativecommons.org/licenses/by/4.0/)

In addition to QKD, quantum teleportation has emerged as a subject of considerable interest, leveraging quantum entanglement to transfer delicate quantum information in a manner that is effectively immune to hacking attempts [579, 862, 863]. This approach has laid the groundwork for the development of quantum networks capable of interconnecting diverse quantum devices, offering unprecedented capabilities that are demonstrably beyond the reach of classical information techniques alone [350, 864]. Quantum secure direct communication (QSDC) [865–867], another significant facet of QCOMM, has also opened avenues for secure data transmission. This methodology has seen rapid advancement in recent years [868–874], empowering users to directly relay sensitive information across secure quantum channels without the need for shared encryption keys. Notably, a QSDC network has been successfully demonstrated with 15 clients [873], showcasing its practical feasibility. Furthermore, by integrating post-quantum cryptography, it is feasible to construct a QSDC network with end-to-end security using existing technological frameworks [874].

Quantum photonic chips represent an optimal foundation for the next wave of quantum technology innovations [73]. Following years of dedicated research, the integration of photonics has been achieved across all components of individual QCOMM systems, encompassing photon sources, encoding and decoding photonic circuits, and detectors [73, 875]. Integrated photonic chips possess the inherent capability to amalgamate numerous advantageous features essential for QCOMM applications, including efficiency, cost-effectiveness, scalability, flexibility, and performance. Moreover, the utilization of wafer-scale fabrication processes further enhances the appeal of chip-based QCOMM systems as a promising platform for the advancement of future quantum technologies [827].

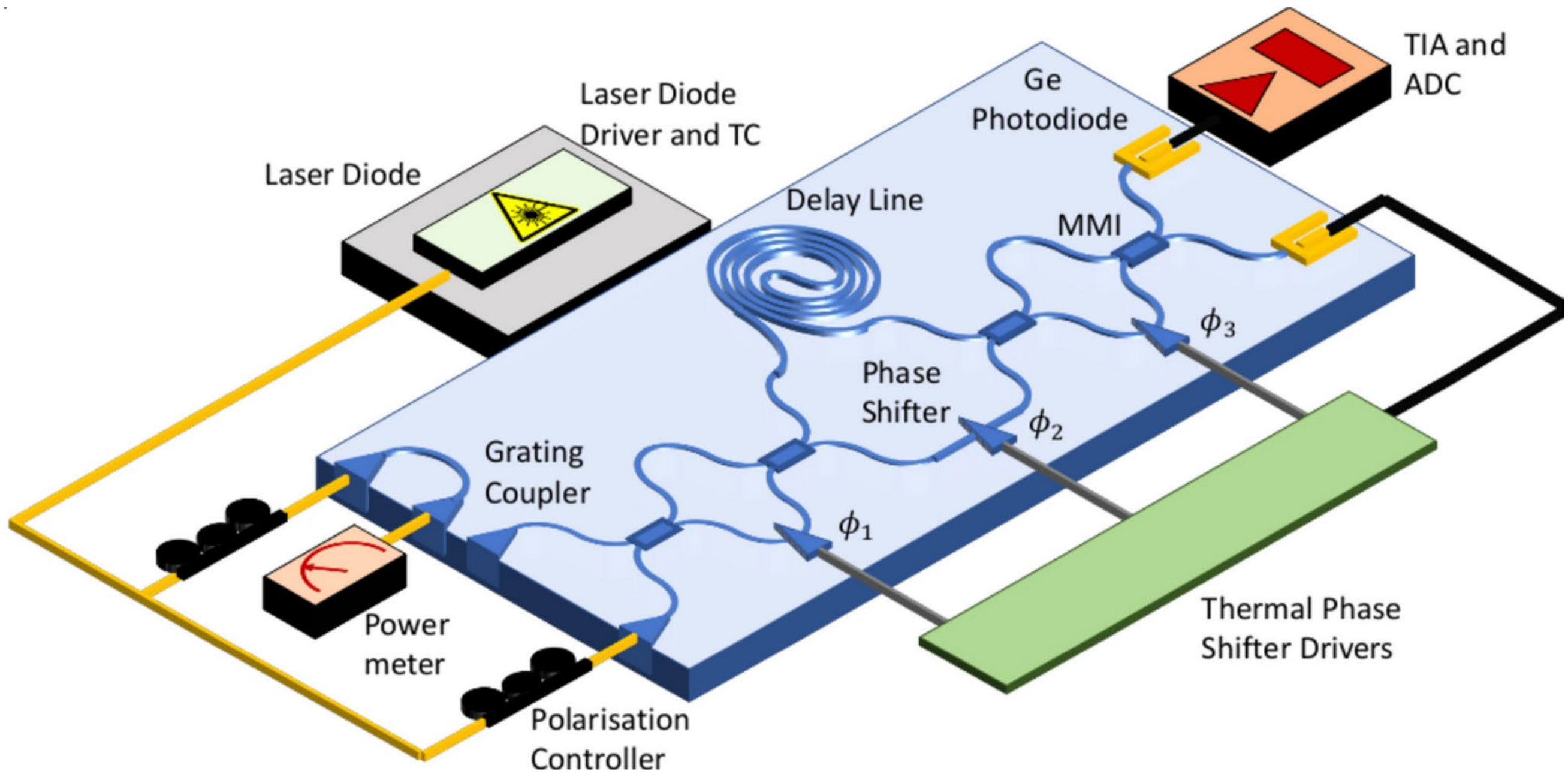

**Fig. 12** Experimental setup overview of integrated QRNG. The experiment employs a diode laser, regulated by a laser diode driver and temperature stabilized via a temperature controller, operating slightly above threshold. A fraction of the light is directed to a polarization controller and introduced into a test waveguide to monitor coupling losses. The remaining light is routed through another polarization controller and subsequently injected into a sequence of MZIs. The first and last MZIs act as tunable B-Ss, while an intermediate unbalanced MZI converts phase fluctuations into intensity fluctuations. At the output of the MZI cascade, two photodiodes are positioned. One serves as a monitoring device to calibrate interferometer phases using heater drivers, which adjust MZI phases through voltage applied to integrated PSs. The second photodiode is linked to a transimpedance amplifier, converting intensity fluctuations into voltage variations. These voltage fluctuations are digitized by an oscilloscope to generate random bits. Reproduced from [1033] under a Creative Commons license (https://creativecommons.org/licenses/by/4.0/)

## 9.3 Implementations of photonic QKD

Quantum communications endeavors to establish secure connections between remote quantum processors. QKD operates by facilitating the creation of a mutually agreed-upon random secret key between two entities. Its notable advantage lies in its capacity to detect potential intrusions by malicious third parties, as any intervention from such actors perturbs the shared system. Subsequent to the establishment of a secret key, the involved parties may commence conventional classical communication protocols [39].

The implementation of QKD predominantly revolves around the utilization of several established protocols, notably the BB84 [273] and E91 [274] schemes, alongside techniques such as third-man quantum cryptography [876] and quantum secret sharing [877]. However, a diverse array of alternative methodologies has also emerged [25, 878, 879], drawing upon DV, CV, or distributed phase reference coding [827]. A fiber-based twin-field QKD (TF-QKD) [880] using sending-or-not-sending (SNS) [881] protocol was demonstrated in [882], over a record distance of 1,002 km. Various QKD implementations is listed in Table 1.

DV-QKD, exemplified by the decoy-state BB84 protocol [828–830], is a notable QKD approach. Additionally, QKD protocols have been proposed, such as CV-QKD, which encodes key information into parameters like the quadrature components of the quantized electromagnetic field [133, 883–885]. Notably, the employment of single photons as carriers of encoded information holds significant promise due to their capacity for entanglement-based secret sharing. Specific protocols such as third-man quantum cryptography and quantum secret sharing hinge upon the utilization of three-particle polarization entangled states $\psi = (|000\rangle + |111\rangle)/\sqrt{2}$, referred to as Greenberger-Horne-Zeilinger or GHZ states [876, 886, 887]. Additionally, schemes employing attenuated lasers have garnered attention, particularly in applications such as coherent one-way (COW) [91, 888–890], differential phase shift (DPS) [879, 892], and decoy-state protocols [893–896]. For a comprehensive understanding of these developments, we refer to [853, 897, 898]. Additional reviews of QKD protocols can be found in [153, 154, 282, 853].

## 9.4 Long-distance quantum communication

Quantum communications hold the potential to facilitate the distribution of quantum information protocols across considerable geographical distances. Upcoming quantum

**Table 1** Quantum key distribution implementations (non-exhaustive list). The degree of integration varies depending on the specific design and technological advancements [827]

| Reference | Year | Protocol[a] | Platform[b] | Encoding | Decoding |
|---|---|---|---|---|---|
| Ma et al. [184] | 2016 | BB84 | Si | ✓ | × |
| Sibson et al. [91] | 2017 | BB84/DPS/COW | InP, $SiO_xN_y$ | ✓ | ✓ |
| Sibson et al. [888] | 2017 | COW/BB84 | Si, $SiO_xN_y$ | ✓ | ✓ |
| Ding et al. [852] | 2017 | HD-QKD | Si | ✓ | ✓ |
| Bunandar et al. [899] | 2018 | BB84 | Si | ✓ | × |
| Paraïso et al. [900] | 2019 | DPS/BB8 | InP | ✓ | × |
| Geng et al. [901] | 2019 | BB84 | Si | ✓ | ✓ |
| Zhang et al. [842] | 2019 | CV-QKD | Si | ✓ | ✓ |
| Dai et al. [890] | 2020 | COW/DPS | Si | ✓ | ✓ |
| Wei et al. [843] | 2020 | MDI-QKD | Si | ✓ | × |
| Cao et al. [844] | 2020 | MDI-QKD | Si | ✓ | ✓ |
| Semenenko et al. [831] | 2020 | MDI-QKD | InP | ✓ | × |
| Paraïso et al. [902] | 2021 | BB84/Modified | InP, Si | ✓ | ✓ |
| Avesani et al. [903] | 2021 | BB84 | Si | ✓ | × |
| Zheng et al. [904] | 2021 | MDI-QKD | Si | × | ✓ |
| Sax et al. [891] | 2023 | BB84 | SG25PIC SiGe | ✓ | ✓ |

[a]DPS differential phase shift, COW coherent one way, HD-QKD high-dimensional QKD, CV-QKD: continuous variable QKD.

[b]Si: Silicon, InP:indium phosphide, $SiO_xN_y$: silicon oxynitride [73, 875, 905]

networks aim to establish the transport of quantum information between remote nodes on a global scale. Noteworthy progress in addressing this challenge is evident in initial demonstrations of first quantum networks conducted in different locations, as outlined in Table 2, including Austria [854], China [906], Japan [857, 907], Switzerland [855], and the USA [908]. Various experimental implementations of QCOMM have been documented [41, 42, 363, 852, 869, 888, 892, 893, 895, 896, 909–928], spanning from targeted applications to the facilitation of communication over extended distances.

QKD [273, 274] holds promise for facilitating secure communication and information transfer [155]. In laboratory settings, the feasibility of point-to-point QKD was initially demonstrated over short distances, starting from 32 cms [850]. Subsequent advancements extended this distance to 100 kms using decoy-state QKD [929, 930], and more recently, to 500 kms with measurement-device-independent QKD [290, 913, 931, 932]. Outside laboratory environments, several small-scale QKD networks have also been successfully tested [854, 857, 858, 908]. However, achieving a global QKD network necessitates a practically secure and reliable infrastructure capable of serving a large number of users spread across extensive geographic areas [933].

The Beijing-Shanghai Backbone Network (B-SBN) [103] represents the forefront of QKD networks globally. Developed by the University of Science and Technology of China (USTC), it stands as the world's pioneering long-range communication link secured by quantum technology [103]. This network, operational in China, interconnects Beijing, Jinan, Hefei, and Shanghai through over 700 fiber links and two high-speed free-space links with the *Micius* QCOMM satellite, orbiting approximately 500 km above Earth [934]. Spanning approximately 2000 km of fiber optic cable among the cities, plus a 2600 km satellite link from Beijing's observatories to near Kazakhstan's border, the B-SBN incorporates 32 trusted nodes crucial for relaying quantum information. These nodes extend in various directions to end users, establishing a robust and secure QCOMM infrastructure [103].

Other prominent contributors to the progression of QCOMM technologies and the establishment of quantum networks, including in-fiber photonic networks, and the development of quantum-safe encryption solutions, encompass: Nippon Telegraph and Telephone Corporation (NTT) [935], University of Geneva [936], and ID Quantique [937]. Toshiba has also played a pivotal role in advancing QCOMM technologies, particularly in in-fiber photonic networks, achieving significant milestones in QKD over extended distances [880, 902, 938]. In 2021, Toshiba Europe [939] and BT Labs [940] collaborated to establish the United Kingdom's first quantum-secure network in Bristol, spanning a 7-km fiber optic cable and connecting three local institutes [941].

Further notable contributors in this domain include Quantum Xchange [942], QuTech [943], China Mobile [944], and the BNL (Brookhaven National Laboratory) Quantum Network [945]. In mid-2022, Amazon

**Table 2** Examples of global fiber-based photonic quantum communication networks, arranged alphabetically (not exhaustive)

| Country | Network | Description | Nodes | Distance | Source |
|---|---|---|---|---|---|
| Austria | SECOQC[a] | The QKD network in Vienna, designed and implemented by the European project SEcure COmmunication based on Quantum Cryptography (SECOQC) | 10 | 200 km | [854] |
| China | Hierarchical Quantum Network | A hierarchical metropolitan quantum cryptography communication network, that was implemented upon the inner-city commercial telecom fiber cables | 32 | 2000 km | [906] |
| China | Beijing-Shanghai Backbone Network | A unified space-to-ground QCOMM network that combines an extensive fiber network comprising over 700 fiber QKD links with two high-speed satellite-to-ground free-space QKD links. Employing a trusted relay structure, the ground-based fiber network spans over 2000 kms. The satellite-to-ground QKD achieves an average secret-key rate of 47.8 kilobits per second during typical satellite passes-more than 40 times higher than previously achieved rates [41]. By integrating both fiber and free-space QKD links, the QKD network extends to a remote node located more than 2600 kms away, enabling any user within the network to communicate with another over a total distance of up to 4600 kms [103]. | 32 | 4600 km | [103] |
| Japan | Tokyo QKD Network | A metropolitan quantum secure communication network, utilizing six integrated QKD systems in a mesh configuration, enables secure TV conferencing over 45 km with GHz-clocked links. The network incorporates a stable commercial QKD product, an application interface for secure mobile communication, and showcases features like eavesdropper detection, secure path rerouting, and key relay through trusted nodes | 5 | 45 km | [857, 907] |
| Switzerland | SwissQuantum | The SwissQuantum QKD network was deployed in the Geneva metropolitan area, operating continuously for over eighteen months. The primary objective of this undertaking was to assess the durability of the quantum layer over an extended duration within an operational setting | 3 | 35 km | [855] |
| USA | DARPA[b] | The DARPA Quantum Network represents the inaugural quantum cryptography network globally, potentially marking the first continuous operation of QKD systems spanning a metropolitan area. This network accommodates diverse QKD technologies, incorporating phase-modulated lasers over fiber, entanglement through fiber, and free-space QKD | 10 | 29 km | [908] |

[a]Secure Communication based on Quantum Cryptography. [b]Defense Advanced Research Projects Agency

announced the "AWS Center for Quantum Networking," aimed at developing innovative hardware, software, and applications for quantum networks [946]. These entities have made substantial advancements in enhancing the capabilities of in-fiber photonic quantum networks and expanding the frontiers of secure QCOMM across extensive distances.

## 9.5 QKD performance parameters

One of the primary challenges confronting QKD technology is its limited operational range, primarily dictated by the transmission of photons through either optical fibers or free space. Both methods encounter significant hurdles over long distances: optical fibers suffer from high absorption losses within the fiber material, while free-space optical transmission sees the beam naturally expand, further constraining transmission capabilities to a few hundred kilometers [947].

To mitigate absorption losses, QKD typically employs wavelength ranges with minimal absorption, known as "telecom windows," located in the infrared spectrum at wavelengths around 1310 or 1550 nm. In contrast, free-space transmission offers lower losses, enabling potential satellite-based communications spanning several thousand kilometers

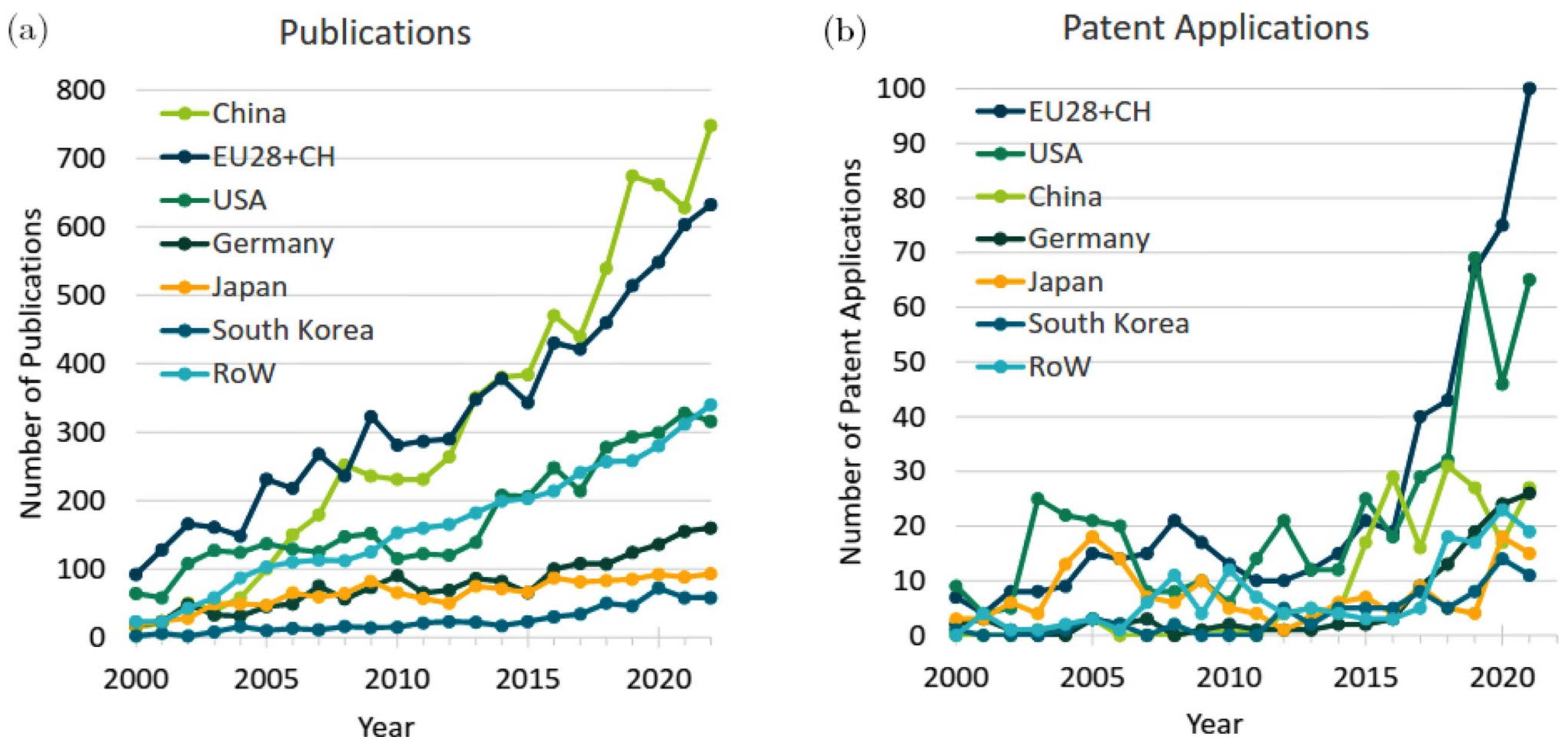

**Fig. 13** Comparative illustration showing the evolution of quantum communication (QCOMM) research outputs and patenting activities from 2000 to 2022. **a** The growth in peer-reviewed publications from 2000 to 2022, focusing on countries with the highest publication outputs. **b** The trends in transnational patent applications in QCOMM from 2000 to 2021, emphasizing the countries with the highest patenting activities. Reproduced from [947] under Creative common license (https://creativecommons.org/licenses/by/2.0/)

using wavelengths in the near-infrared range, approximately 800 to 850 nm [948].

Direct QKD connections typically span distances ranging from 100 to a maximum of 400 or 500 km. These distances are constrained primarily by losses within optical fibers, where transmission typically drops to around 1% after 100 km, as well as by the inherent dark noise of the detectors used. The longest direct connection achieved to date using optical fiber spans 1000 km, employing twin-field QKD, which utilizes the interference of two phase-stable optical fields to exchange quantum information between communication partners [882].

Additionally, the transmission clock rate plays a critical role in QKD implementation. Currently, DV-QKD achieves key rates of 110 Mbps over 10 km of optical fiber [161, 662, 882], while CV-QKD systems have achieved key rates ranging from 49 Mbps over 25 km to 2 Mbps over 80 km of fiber optic cable [949]. For a comprehensive insights of current performance parameters, Zhang et al. provide a comprehensive overview, particularly focusing on CV-QKD [950] (see Fig. 1 in [103]).

### 9.6 Emerging trends in quantum communication

The field of QCOMM is predominantly shaped by research and development efforts from industry and academic institutions. Over the past two decades, there has been a substantial and consistent increase in the volume of research literature pertaining to QCOMM (see Fig. 13a). In the year 2000, approximately 200 publications globally addressed this topic, a number that surged to nearly 2000 by 2022 [947]. Notably, the majority of these publications originate from authors affiliated with institutions in China, closely followed by those from the European Union (EU28 + CH), and approximately half as many from the United States. Japan and South Korea also contribute significantly, along with Canada, India, Russia, Australia, and Singapore, collectively categorized as "rest of world—RoW" in Fig. 13.

Additionally, Fig. 13b depicts the dynamics of QCOMM-related patent applications from various countries and the EU between 2000 and 2021. It illustrates a robust increase in patenting activities, rising from fewer than 20 applications in 2000 to over 200 in 2021, with notable acceleration since 2014 [947].

## 10 Quantum computational advantage with photons

### 10.1 Regime of quantum computational advantage

In the pursuit of quantum supremacy [951, 952], extensive efforts have been invested in advancing quantum computers to perform tasks beyond classical computing capabilities [1]. This regime, also known as quantum computational advantage, represents the critical juncture where a quantum computer demonstrates the ability to surpass classical systems. Two prominent quantum computing frameworks for achieving this objective are BS [556] and GBS [773]. These paradigms leverage quantum interference among indistinguishable particles, typically bosons, to execute specialized computational operations.

The progress of photonic quantum computing is intricately linked with the advancement of BS and GBS algorithms, owing to certain technical disparities between photonic quantum computers and other quantum computing platforms. Although it is anticipated that a fully developed photonic quantum computer will possess universal quantum computation capabilities, the majority of experimental endeavors within the realm of photonic platforms have thus far centered on variants of BS and GBS [953]. Remarkably, even with the sole application of BS, photonic quantum computing has emerged as one of the pioneering platforms to assert quantum supremacy [1]. Notably, integrated photonic chips find versatile utility in artificial neural networks [842, 954] and QKD [153, 842].

## 10.2 Boson sampling

Boson sampling (BS) has emerged as a valuable instrument for investigating quantum advantages [1] over classical computing, notably as it obviates the need for universal control over the quantum system. This attribute aligns with the capabilities inherent in existing photonic experimental platforms. The significance of BS [556] is underscored by its ability to captivate both theorists and experimentalists, offering insights into the superior computational prowess of quantum systems while challenging the Extended Church-Turing thesis (ECTT), all without necessitating the full capabilities of a universal quantum computer.

In the BS paradigm, indistinguishable photons traverse an array of optical elements, engendering intricate interference patterns, with the resulting output distribution encoding information pertaining to the underlying quantum states and network parameters. The primary objective is to sample from this distribution, a computational task posited to be heavy for classical computers, thereby illustrating QCOA [556]. Nevertheless, the initial BS approach encounters substantial impediments related to scalability and experimental implementation, primarily attributable to the stringent requirement for a dependable source of numerous indistinguishable photons.

BS was initially posited as a prospective quantum computational supremacy benchmark by Aaronson and Arkhipov in 2011. Aaronson and Arkhipov posited that the efficient simulation of a passive linear optics interferometer, utilizing single-photon state inputs, is inherently challenging [556]. This model stands as a non-universal quantum computing paradigm that is considerably more feasible to construct than universal quantum computing. In the BS framework, an ensemble of $n$ indistinguishable bosons is directed into an $m$-mode interferometer, characterized by Haar-random transformations. The output distribution, sampled in the photon number basis, relies on the intricate ties between probability amplitudes and the calculation of permanents of submatrices-a computational problem acknowledged to be #P-complete due to the statistical properties of bosons.

The experimental setup involves introducing $n$ photons into an optical circuit comprising $m$ modes, where $m$ significantly exceeds $n$. This state undergoes manipulation through a sequence of PSs and B-Ss. The PS introduces a phase $R(\varpi) = \mathrm{e}^{\mathrm{i}\varpi_l}$, where $\varpi_l$ denotes an angle, exclusively to the amplitude in mode $l$, while maintaining identity in the remaining $m-1$ modes. While, a B-S, acting on two modes with a rotation matrix $\begin{pmatrix} \cos\chi & -\sin\chi \\ \sin\chi & \cos\chi \end{pmatrix}$, where $\chi$ is an angle, functions as identity in the remaining $m-2$ modes. Ultimately, a measurement is conducted to ascertain the number of photons present in each mode. Figure 5 illustrates an optical circuit encompassing these elements. Each observed measurement outcome serves as a sample from the symmetric wavefunction inherent to bosonic systems.

Aaronson and Arkhipov discerned that the presence of an efficient classical algorithm for sampling from this distribution implies the existence of a classically efficient algorithm for calculating the permanent of an associated matrix. The realization of such an algorithm would precipitate the collapse of the polynomial hierarchy to the third order, an eventuality deemed improbable according to the conjectures posited by in [955]. Consequently, the existence of such an algorithm is deemed unlikely.

Despite its robust theoretical promise, BS itself has encountered significant challenges preventing widespread practical applications beyond laboratory settings. These challenges primarily revolve around the intricate control required over photon interactions and the scalability of experimental setups, major challenges are mode-mismatch, photon loss, single photon state preparation and detection imperfections and network errors [956]. Achieving precise manipulation of indistinguishable photons at scale remains technical hurdle, crucial for reliable and accurate BS experiments [957–961]. Nevertheless, the foundational work by Aaronson and Arkhipov has established BS as a compelling candidate for demonstrating quantum advantage and challenging classical computational capabilities.

## 10.3 Gaussian boson sampling

In response to the experimental limitations posed by traditional BS, a pioneering methodology known as Gaussian boson sampling was introduced in 2017 [773]. GBS strategically addresses computationally challenging problems by leveraging squeezed states as a non-classical resource. This innovative approach aims to elucidate the intricacies of sampling from a general squeezed state, introducing a novel expression that establishes a connection between the probability of measuring a specific photon output pattern from a general Gaussian state and the *Hafnian* matrix function [773]. GBS extends its applicability to CV quantum states and operations, employing Gaussian states and transformations to handle variables such as position and momentum while adhering to the foundational principles of interference.

GBS represents a modification of the BS paradigm, wherein Gaussian states, as opposed to photon states, serve as inputs to the optical circuit [773]. Gaussian states are characterized by Wigner quasi-probability distributions (WQDF) [962, 963], exhibiting a Gaussian form. Notably, Gaussian states offer the advantage of deterministic generation [773, 964] and introduce additional degrees of freedom relative to traditional boson sampling. While BS corresponds to sampling from the permanent of a matrix, GBS is computationally equivalent to sampling from the *Hafnian* function of a matrix.

In the context of a graph $\mathcal{G}$ with an adjacency matrix $\Sigma$, the *Hafnian* of $E$ represents the count of perfect matchings in the graph $\mathcal{G}$. A matching in a graph is a subset of edges denoted as $\mathcal{M}$, such that no two edges in $\mathcal{M}$ share a common vertex. A matching $\mathcal{M}$ is classified as perfect if each vertex is incident to precisely one edge in $\mathcal{M}$. In contrast, the Permanent yields the count of perfect matching exclusively for bipartite graphs. Consequently, the *Hafnian* can be conceptualized as a generalization of the Permanent. The relationship between the *Hafnian* and the Permanent, utilizing the adjacency matrix $\Sigma$, is expressed as:

$$\mathit{Hafnian}\begin{pmatrix} 0 & \Sigma \\ \Sigma^T & 0 \end{pmatrix} = \text{Permanent}\,(\Sigma).$$

Both BS and GBS stand as pivotal experiments in the pursuit of quantum supremacy, showcasing the potential computational advantages of quantum systems in specific, albeit challenging, computational scenarios. For a more comprehensive exploration of GBS interested readers are encouraged to consult [964]. Additionally, [119, 513, 514, 965–969] for a deeper understanding of GBS applications.

## 10.4 Achieving quantum computational advantage with photons

Exploration into computing beyond the classical realm was undertaken by Google Quantum AI and Collaborators [970, 971] and Xanadu [45], utilizing superconducting and photonics quantum computers, respectively. Additionally, researchers in China [385] have also showcased similar experiments using both photonic [43, 108] and superconducting architectures [972, 973]. For a detailed timeline of these experiments, refer to [1]. The objective of these experiments was to sample from the output distribution of random quantum circuits [974, 975].

Several companies have commenced the commercialization of photonic quantum computing technology. For instance, Xanadu has developed a comprehensive hardware-software system designed for integrated photonic chips. These chips have the capacity to execute algorithms that necessitate up to eight modes of squeezed vacuum states, initialized as two-mode squeezed states within a single temporal mode, in conjunction with a programmable four-mode interferometer [100]. Although their initial photonic chip is comparably smaller in scale when set side by side with larger photonic platforms, such as those utilized in extensive GBS experiments as seen in [1, 43], Xanadu posits that this platform exhibits promising scalability, especially in light of recent advancements in chip manufacturing technologies. Importantly, it retains its dynamic programmability. Notably, Xanadu has also unveiled a photonic processor capable of executing a GBS experiment utilizing 216 squeezed modes, facilitated by a time-multiplexed architecture [45].

A small-scale BS experiments was conducted in 2013 [55, 56, 82, 83], showing photon distribution proportional to the square of the permanent modulus. These experiments utilized SPDC [168] sources with inherent probabilistic behavior. To address these limitations, scattershot BS was introduced in 2014 [774] and subsequent demonstrations in [93, 976]. Despite theoretical elegance, experimental realization faces challenges due to ultra-high heralding efficiency, rapid optical switches, and surplus SPDC sources. A more direct solution involves on-demand single photon sources driven by a quantum two-level system. In 2017, Wang et al. achieved a BS experiment with 5-photons, surpassing prior efforts by 24, 000 times [95]. In 2019, BS with 20 input photons demonstrated an output state space exceeding $3.7 \times 10^{14}$ dimensions by over 10 orders of magnitude [99].

GBS offers a more efficient path for quantum computational advantage. In a 2020 experiment by Zhong et al. [43], GBS demonstrated quantum advantage. Enhanced with 50 single-mode squeezed-state inputs and a 144-mode

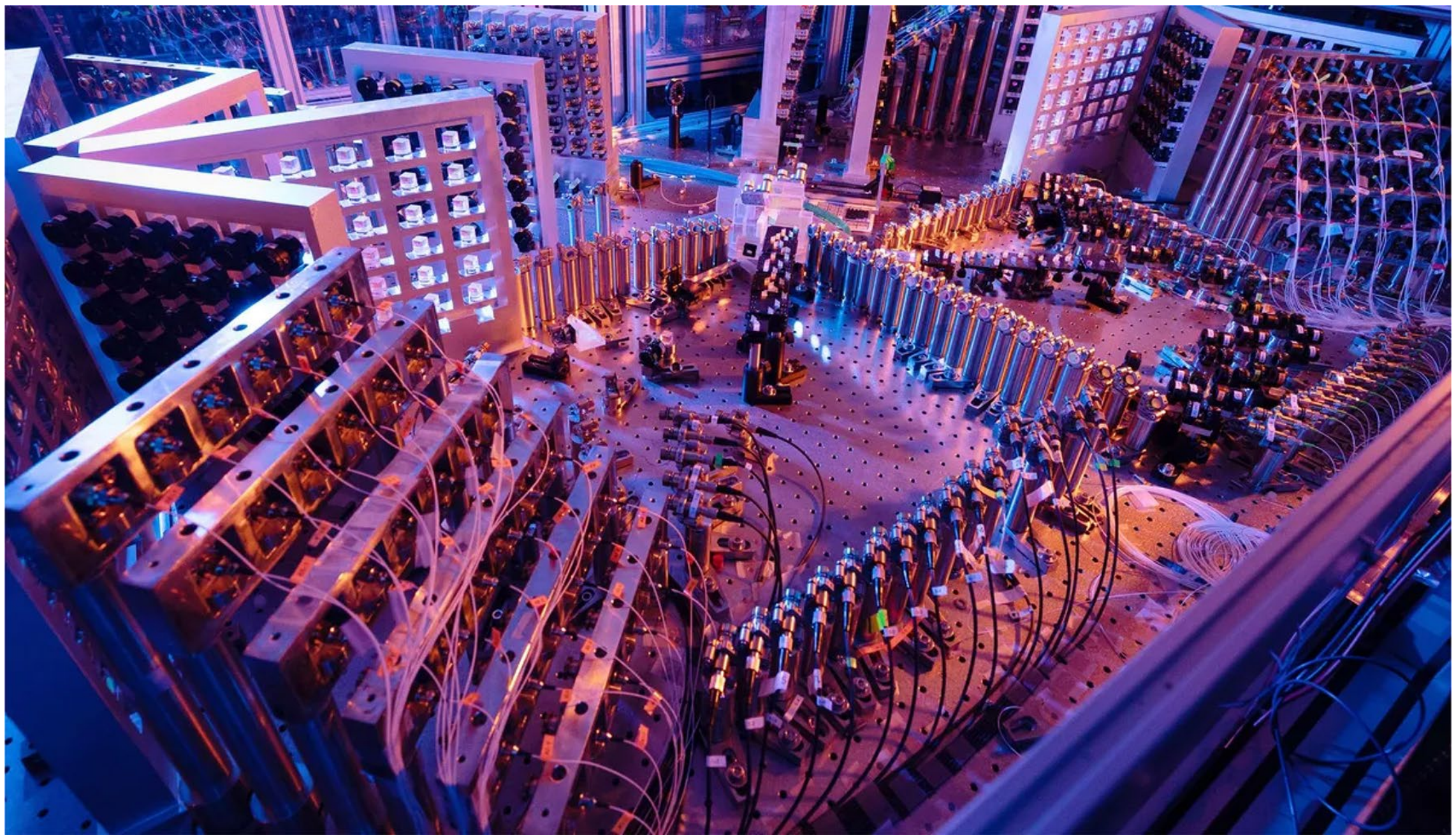

**Fig. 14** The *Jiuzhang* light-based quantum computing device, engineered by USTC. The machine operates by intricately manipulating light through an array of optical components. This visual representation of the Jiuzhang photonic network provides insight into its experimental configuration, which occupies an optical table spanning an area of approximately three square meters. Within this setup, 25 Two-Mode Squeezed States (TMSSs) are introduced into the photonic network, resulting in the acquisition of 25 phase-locked light signals. To provide further elucidation, the output modes of the *Jiuzhang* photonic network are systematically segregated into 100 distinct spatial modes through the employment of miniature mirrors and polarizing B-Ss. This accomplishment signifies the emergence of the second quantum computing system to assert the achievement of quantum computational advantages, following Google's *Sycamore* quantum processor [21, 22, 970]. Reproduced under a Creative Commons license (https://creativecommons.org/licenses/by/4.0/) from [43]

interferometer. Named *Jiuzhang* [44] (see Fig. 14), these photonic quantum computers exhibit a state space of $10^{43}$ and a sampling rate $10^{10}$ times faster than current supercomputers. *Jiuzhang* is partially programmable through precise control of input Two-Mode Squeezed States (TMSSs). Time-bin-encoded BS, introduced by He et al. in 2017 [237], provides full programmability. Combining GBS and time-bin loops, Madsen et al. [45] demonstrated quantum advantage with 216 inputs, 16 photon-number-resolving detectors, and up to 219 detected photons. The output probability of GBS, linked to the *Hafnian* and perfect matchings in a graph, suggests practical applications [53, 513, 514, 966]. GBS is poised to become a specialized photonic platform for real-world applications [1], advancing toward NISQ processing [977]. The computational complexity of simulating a noisy rendition of GBS has been explored [978], and GBS has recently emerged as the second platform to demonstrate quantum computational supremacy [43]. A pioneering experimental effort in the realm of dynamically programmable GBS nanophotonic chips was conducted by [100].

In 2023, a new GBS experiment employing pseudo-photon-number-resolving detection was detailed by Deng et al. in [108], recording photon-click events of up to 255, as illustrated in Fig. 15. The investigation incorporates considerations for partial photon distinguishability and advances a comprehensive model for characterizing noisy GBS. Within the realm of QCOA, Bayesian tests and correlation function analysis are employed to authenticate the samples against existing classical spoofing simulations. Comparative estimations with the most advanced classical algorithms indicate that generating a single ideal sample from the same distribution on the supercomputer *Frontier* would necessitate approximately 600 years using exact methods, while the quantum computer, *Jiuzhang 3.0*, accomplishes this task in only 1.27 microseconds. The generation of the most challenging sample from the experiment using an exact algorithm would require *Frontier* approximately $3.1 \times 10^{10}$ years.

## 11 Applications of photonic quantum computers

Leveraging the inherent quantum characteristics of qubits [979], quantum algorithms can be formulated by harnessing novel quantum-parallel and entangled properties. With recent advancements in commercial quantum computing,

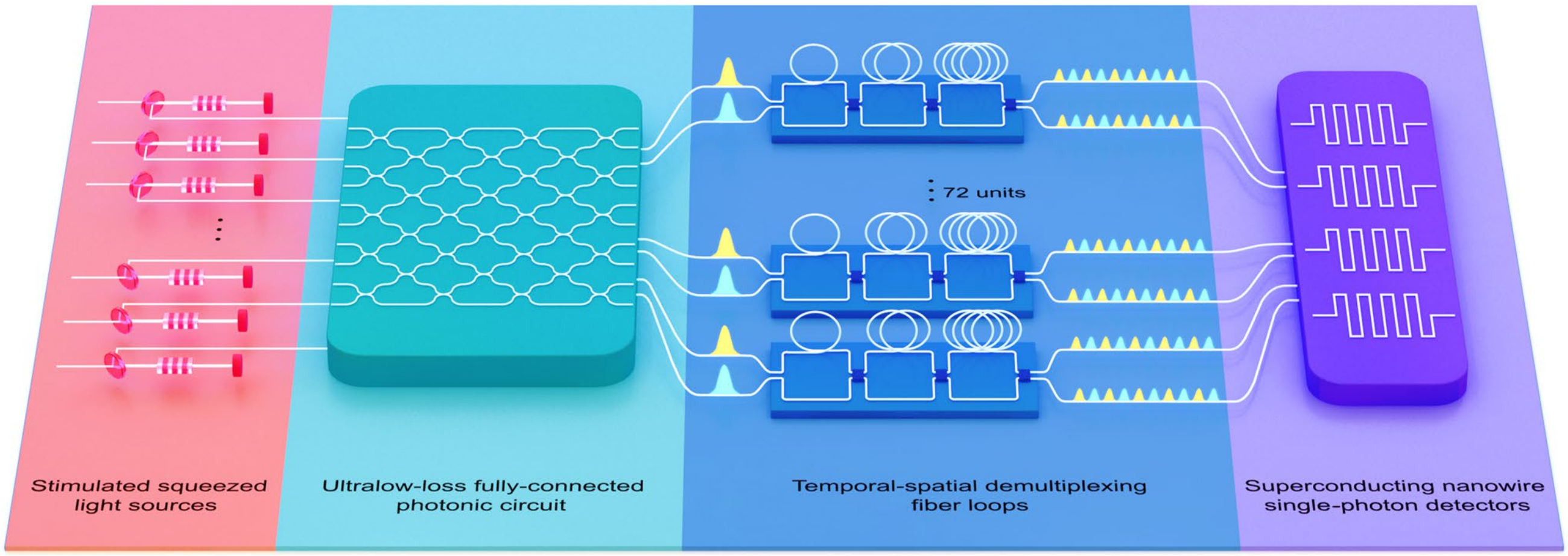

**Fig. 15** The experimental configuration entails 25 stimulated two-mode squeezed state photon sources synchronized in phase and directed into a 144-mode ultralow-loss optical interferometer with full connectivity. These photons traverse 72 units of fiber loop setups for temporal-spatial demultiplexing before being captured by 144 superconducting nanowire single-photon detectors, constituting a pseudo-photon-number resolving detection system. Each fiber loop setup incorporates two input modes denoted by distinct colors. Photon streams from each mode undergo temporal demultiplexing via fiber B-Ss and delay lines, dividing them into four time bins, with each bin further divided into two path bins at the final fiber beam splitter. The photons originating from the same fiber loop setup's two input modes can be distinguished based on their temporal bin parity through a coincidence event analyzer (not depicted). Regenerated under a Creative Commons licenses (https://creativecommons.org/licenses/by/4.0) from [108]

the anticipated impacts of quantum computations span diverse fields [1], including but not limited to quantum machine learning (QML) [980–985], chemistry [986–989], drug discovery [990], quantum sensing [991], cryptography [41, 155, 857, 897, 913, 924, 925, 932, 992, 993], and combinatorial optimization [994]. An essential application of quantum computation lies in quantum systems' simulation [986–988], accelerating the research and development processes for materials and drugs [995]. Various quantum algorithms, such as QPE (quantum phase estimation) [996] and VQE (variational quantum eigensolver) [89, 997–1000], have been devised to compute the ground state energy of molecules [89]. The foundational subroutine within the QPE algorithm is the QFT (quantum Fourier transformation) [25].

Combinatorial optimization, pervasive in everyday life, presents challenges for classical computers to efficiently solve certain NP-hard problems, such as the traveling salesman problems (TSPs), within polynomial time [1001]. The advent of quantum computations holds promise as an alternative solution to these combinatorial problems [1002–1004]. By casting problem formulations into the Ising model and quadratic unconstrained binary optimization (QUBO) [1005], they can be embedded in the quantum processor's graph. Quantum annealing [1006], an algorithm addressing optimization problems, has shown proof-of-concept success on various problems, including TSP and the nurse scheduling problem [1007].

The coherent Ising machine (CIM) excels in solving Ising-like optimization problems, while the photonic GBS machine is proficient in computational challenges such as dense graph optimization. Ising problems and dense graph problems, known for their NP-hard nature [1008], find effective resolution solely through photonic quantum systems. Programmable universal photonic quantum computers face challenges of scalability and reliability, akin to other physical approaches. This underscores the necessity for a fault-tolerant universal computer as the long-term objective.

Quantum computing utilizing integrated photonic chips has garnered significant attention in recent years. Two distinct optical models have emerged [1009]: specific quantum computing models, exemplified by BS (e.g., models [961] and [556]), and universal quantum computing models (UQCM) [979], such as one-way or MBQC [8, 16, 18, 62, 1010, 1011]. In the realm of specific quantum computation, various photonic systems have been successfully demonstrated through the utilization of quantum photonic chips [55, 56, 58, 82, 83, 93, 95, 100, 773], enabling a seamless and efficient implementation of BS.

GBS [773, 964], an advancement capable of significantly enhancing the sampling rate through the integration of squeezed light sources, has been executed for the computation of molecular vibronic spectra on both Si [58] and SiN [100] chips, accommodating up to 8 and 18 photons, respectively. Recently, QCOA regime has been achieved by employing photonic GBS processors [43, 45], opening avenues for the further advancement of integrated specific quantum computers. These systems hold promise for diverse applications, including complex molecular spectra analysis [1012], graph optimization [514], quantum chemistry [1013], molecular docking [966], among others.

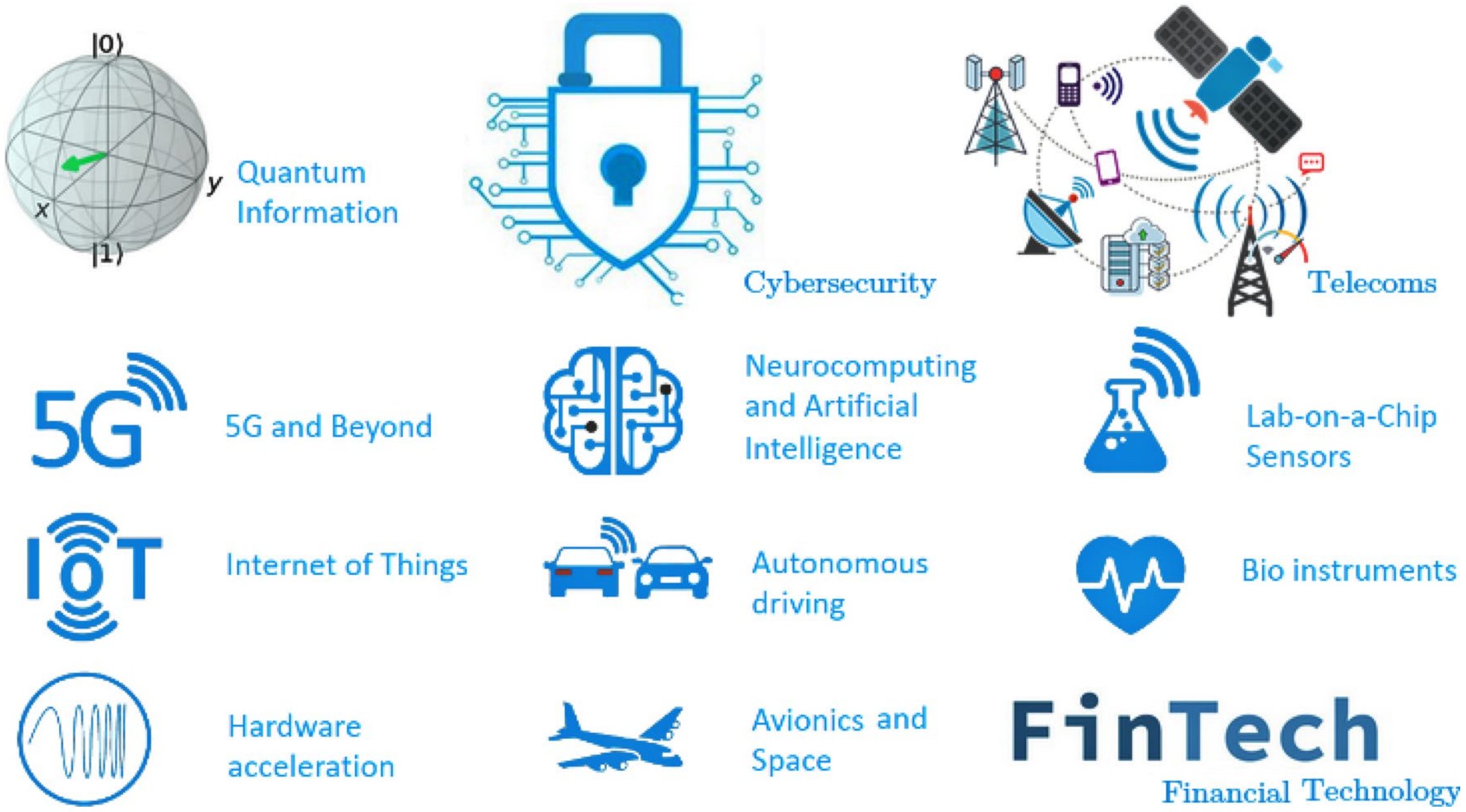

**Fig. 16** Examples of emerging applications demanding heightened processing speed capabilities. Adapted from [811]

Concerning UQCM [204, 1014, 1015], on-chip photonic components have demonstrated crucial functionalities, such as the CNOT gates [46, 118], as well as the implementation of compiled Shor's factorization [75]. Additionally, concerted efforts in both architectural and technological domains have been dedicated to photonic one-way quantum computation. This approach, relying on cluster states and sequential single-qubit measurement, facilitates the execution of universal quantum algorithms [8, 16, 157], with notable enhancements achievable through the native implementation of resource state generation and fusion operations [789, 1016, 1017]. The circuit implementations aligned with this approach include programmable four-photon graph states on a Si chip [98], programmable eight-qubit graph states on a Si chip [1018], and path-polarization hyperentangled and cluster states on a $SiO_2$ chip [94].

The inherent characteristic of high re-programmability stands as a crucial aspect of IQPs, facilitating the versatile processing of tasks. The progress in quantum hardware development on IQP chips, encompassing both gate-based and MBQC approaches for QIP, along with their respective algorithmic implementations [73]. These advancements include chip-based QCOMMs [153, 154, 281, 282, 827, 853, 899], QRNGs [1019] via different schemes [749, 1020–1033], gate-based QIP [46, 75, 172, 841, 1034], programmable quantum chips for multifunctional QIP [47, 49, 53, 57, 89, 1035], entanglement generation, manipulation and measurement and MBQC [94, 97, 98, 120, 167, 766, 1036], on-chip sampling of photons and quantum simulation [33, 989], multiphoton interference [1037, 1038], validation of BS [46, 84, 1039–1041], regime of quantum advantage [33, 1042], simulation via quantum walks [77, 85, 1043–1049], and molecular simulation [53, 58, 89, 96, 513, 989, 996, 1050].

In the past ten years, significant progress in the field of photonic quantum technologies has resulted in increased complexity of systems, leading to breakthroughs in various aspects of quantum information science. Notable accomplishments encompass the realization of quantum advantage [43–45] and the establishment of satellite QCOMMs [41, 42]. Recently, photonic processors, have garnered increasing attention due to their diverse range of applications. See Fig. 16. These applications extend to QIP utilizing linear optics [46–58], QML [59–63], quantum repeater networks [66–69], and radio-frequency signal processing [64, 65]. Quantum photonic chips have undergone rapid maturation, evolving into a versatile platform of considerable significance in the advancement of state-of-the-art QCOMM technologies. For recent advances in quantum photonic chips for communication and internet, see [827]. For near-term photonic quantum computing applications, including software and algorithms, refer to [119]. For an overview of integrated photonic quantum technologies, consult [73]. For a detailed review of applied quantum computing, including proof-of-concept experiments, see [1].

## 12 Prospects, implications, and challenges in photonic quantum computing

Looking ahead, the prospects for photonic quantum computing are compelling. Researchers envision a future where photonic quantum computers seamlessly integrate into existing communication networks, offering secure and ultra-fast data transmission alongside potent computational capabilities. Industries reliant on intensive computational tasks, such as pharmaceuticals for drug discovery and logistics for optimization, could benefit immensely from the quantum speedup enabled by photonic processors.

The emergence of quantum computing has ignited a compelling journey in the development of optical quantum technology, presenting an intriguing realm for exploration and a crucial practical pursuit. Photonic quantum technologies show significant promise across a spectrum of applications (see Sect. 11), thanks to the unique advantages inherent in single photons that render them a preferred candidate for quantum information transmission across diverse domains. Recent advancements in technology have now facilitated the realization of tangible applications in photonic QIP.

According to [1051], it is projected that the worldwide photonics market will attain a size of USD 837.8 billion by 2025, with a CAGR (compound annual growth rate) of 7.1% from 2020 to 2025. The significant expansion of photonics-enabled products in healthcare, communication and information technologies, and industrial manufacturing sectors is the primary driver of this global market growth, expected to continue propelling market expansion in the near future.

While photonic quantum computing holds promise for revolutionizing information processing, several challenges must be overcome to fully exploit its potential. Maintaining quantum coherence and minimizing photon loss are critical hurdles that require advancements in photon sources, noise reduction techniques, and error correction protocols. Additionally, integrating diverse photonic components into coherent quantum systems poses engineering challenges, necessitating innovations in fabrication techniques and robust control mechanisms for photonic quantum circuits. Continued research into photon generation, manipulation, and detection will pave the way for more efficient quantum circuits and enhanced control mechanisms. Addressing these obstacles will be pivotal in unlocking the transformative potential of photonic quantum computers and ushering in a new era of QIP.

## 13 Conclusion

In the evolving landscape of quantum technologies, photonic quantum computing emerges as a beacon of promise and innovation. Recent strides in integrated photonics, quantum light sources, and photon manipulation techniques are propelling the evolution of photonic quantum computing. These advancements facilitate enhanced scalability, increased qubit counts, and greater circuit complexity.

The integration of photonics into quantum computing carries profound implications across various domains. Photonic quantum computers possess the potential to exponentially accelerate computational tasks across a spectrum of applications, encompassing cryptography, material science simulations, optimization algorithms, surpassing the capabilities of classical systems. Practical applications extend to secure communication protocols, precise molecular simulations crucial for drug discovery, and the optimization of logistical networks, highlighting their transformative impact on industries that rely heavily on intensive computational tasks. Moreover, photonic quantum computers hold promise for simulating complex quantum phenomena with unprecedented accuracy, advancing fields like quantum chemistry and materials science. Furthermore, their ability to perform complex optimization tasks could revolutionize fields ranging from finance to energy management.

This article aims to furnish readers with a comprehensive overview of this dynamic field. Nevertheless, the significant dedication to advancing photonic quantum computing technologies suggests that additional achievements may have emerged during the completion of this review, highlighting the challenges in maintaining current knowledge in this rapidly evolving field.

**Acknowledgements** The author sincerely thanks iPronics for granting access to their white paper, a valuable resource that significantly enriched this research. The author thanks the anonymous reviewers for their insightful feedback and constructive suggestions, which significantly enriched and clarified this work. The views and conclusions expressed in this work are solely those of the author and do not reflect the official policy or position of The photonic quantum computers, Jiuzhang, Xanadu Quantum Technologies, iPronics Programmable Photonics, University of Science and Technology of China, University of Geneva, ID Quantique, Toshiba, BT Labs, Nippon Telegraph and Telephone Corporation (NTT), Quantum Xchange, QuTech, China Mobile, The Brookhaven National Laboratory, AWS Center for Quantum Networking, Intel, Arqit, Huawei, LG Electronics, Toshiba, QuantumCTek, Deutsche Telekom, British Telecom, Ericsson, IBM, PsiQuantum, Eagle Technology, Corning, Microsoft, Massachusetts Institute of Technology (MIT), Fraunhofer, Delft University of Technology, South China Normal University, or any affiliated organizations.

**Author contributions** M. A.: Conceptualization, Critical Analysis, Literature Review, Resources, Visualization, Investigation, Validation, Writing, Review, and Editing.

**Funding** The author declares that no funding, grants, or any other forms of support were received at any point throughout this research work.

**Availability of data and materials** The datasets generated during and/or analyzed during the current study are included within this article.

The data sets produced and/or analyzed during the current study are incorporated within this article.

## Declarations

**Ethics approval and consent to participate** Not applicable.

**Consent for publication** The author has approved the publication. This research did not involve any human, animal or other participants.

**Competing interests** The author declare that he has no known competing financial interests or personal relationships that could have appeared to influence the work reported in this paper.